\documentclass[preprint2]{aastex}

\shorttitle{Lithium depletion and metallicity  in M35}
\shortauthors{Barrado y Navascue\'es, Deliyannis, \& Stauffer}

\begin{document}

\title{WIYN\footnote{The WIYN telescope is maintained and operated by  a 
consortium whose member institutions are University of 
Wisconsin, Indiana University, Yale University, and the 
National Optical Astronomy Observatories.} Open Cluster Study 5.  Lithium 
Depletion and Metallicity  in G and  K Dwarfs of the Open Cluster M35.}

\author{David Barrado y Navascu\'es\altaffilmark{2}}
\affil{Max-Planck-Institut f\"ur Astronomie. K\"onigstuhl 17,
Heidelberg, D-69117 Germany. barrado@pollux.ft.uam.es}

\author{Constantine P. Deliyannis}
\affil{Astronomy Department, Indiana University,
Swain Hall West 319, 727 E. 3rd Street, 
Bloomington, IN 47405-7105, USA.
con@athena.astro.indiana.edu}      

\author{John R. Stauffer\altaffilmark{3}} 
\affil{Harvard--Smithsonian Center for Astrophysics,
 60 Garden St., Cambridge, MA 02138, USA.
jstauffer@cfa.harvard.edu}

\altaffiltext{2}{Present address: Departamento de F\'{\i}sica Te\'orica, C-XI.
Universidad Aut\'onoma de Madrid, Cantoblanco, E-28049 Madrid, Spain }

\altaffiltext{3}{Present address: IPAC,  California Institute of Technology,
 Pasadena, CA 91125, USA}

%%%%%%%%%%%%%%%%%%%%%%%%%%%%%%%%%%%%%%%%%%%%%%%%%%%
%%%%%%%%%%%%%%%%%%%%%%%%%%%%%%%%%%%%%%%%%%%%%%%%%%%
%%%%%%%%%%%%%%%%%%%%%%%%%%%%%%%%%%%%%%%%%%%%%%%%%%%
\begin{abstract}

We present an analysis of high quality spectra of members of the young
cluster M35. By using a multi-fiber spectrograph, we are able to 
collect high signal--to--noise, high resolution  spectra of a sample
of photometric candidate members.
 Accurate radial velocities are used to establish 
the membership status and rotational velocities are measured using
 cross-correlation. We also derive the metal content 
of the cluster,   [Fe/H]$_{\rm M35}$=--0.21$\pm$0.10,
 based on spectral synthesis.
Finally, we  derive the lithium abundances of the {\it bona fide} 
cluster members and compare the results with members of other clusters.
For example, M35 shows a smaller range in both rotation
rates and lithium abundances as compared to the Pleiades.  We discuss possible
roles of various parameters.
Our high quality M35 database of lithium abundances and rotational velocities
are perfectly suited to be used as a laboratory
to test theoretical models dealing with the lithium depletion phenomenon. 
We discuss the role of stellar inhomogeneities and rotation on the
lithium depletion phenomenon.

\end{abstract}


%%%%%%%%%%%%%%%%%%%%%%%%%%%%%%%%%%%%%%%%%%%%%%%%%%%
%%%%%%%%%%%%%%%%%%%%%%%%%%%%%%%%%%%%%%%%%%%%%%%%%%%
%%%%%%%%%%%%%%%%%%%%%%%%%%%%%%%%%%%%%%%%%%%%%%%%%%%
\section{Introduction}

Late spectral type  stars share a series of characteristics
which make them extremely interesting objects. Among other
parameters, these properties usually depend on the age and
the stellar mass. These stars have activity which is
analog to the one present on the Sun, arising from the stellar
corona, the chromosphere and the  photosphere.  This
activity is produced as a consequence of the interaction  between
differential rotation and magnetic fields, the so-called 
stellar dynamo (Parker 1955). Rotation itself is
 a very important characteristic,
with a dependence on mass and age not fully understood.
 The understanding of these 
properties require to locate them in an evolutionary scenario.
This is done by studying members of open clusters of different
ages and metallicities, so we can determine which is the
particular role of age, stellar mass and chemical composition.

Lithium is a chemical element which can be used as a probe to the
internal structure, since convection
transports the material downward to layers hot enough  to
 destroy lithium
by interactions with protons. Due to this, its surface abundance
depends strongly on age for a given mass during the pre-Main
Sequence (pre-MS) phase. And, because of
increasing depth of the convective envelope when moving to less
massive stars, the lithium abundance also depends on mass.
(or stars of 
spectral type later than F8, at a given age, more massive 
stars have higher Li.)
After the pre-MS phase, additional mixing mechanisms are needed,
in order to explain the lithium abundance of FGK spectral type
members of clusters of different ages.
For recent reviews, see Deliyannis (2000) and Jeffries (2000).
In particular, Deliyannis (2000) stresses  the stellar mass as  
the first parameter controlling Li depletion, 
and age the second parameter.  Additional parameters might be 
composition, and initial angular momentum. 
There have been reported a significant amount of studies on the 
evolution of lithium with age and the dependence on mass.
Updated reviews on the lithium phenomenon,
 from the theoretical and observational point
 of views,  can be found in Vauclair (2000) and Mart\'{\i}n
(2000), respectively. Very well known open clusters such as 
the Hyades,  Praesepe, Coma, M67 and Alpha Per  have been
exhaustively investigated, as well as other not so famous, and
the behavior of lithium in their members analyzed.
The Pleiades has played a {\it prima donna} role in these
investigations. 
Duncan \& Jones (1983) showed that G Pleiades stars have
significant lithium abundances differences.
Butler et al. (1987) reported a scatter for rapid rotators of  K
spectral type, confirmed by 
 Garc\'{\i}a L\'opez et al. (1991a,b),  and
Soderblom et al. (1993a).  Garc\'{\i}a L\'opez, Rebolo, \& Mart\'{\i}n
(1994) added more data to the sample and
established that the relation between fast rotators and high
lithium abundances broke down for stars at Teff$<$4500 K.
Jones et al. (1996) examined the late K and Russell (1996), Stuik
et al. (1997), Jeffries (1999), King, Krishnapurthi \&
Pinsonneault (2000) and Barrado y Navascu\'es et al.
 (2000) have recently  dealt with the origin 
of the lithium spread in K Pleiades stars, without solving
completely the puzzle.

The M35 cluster (NGC2168) is, allegedly,  coeval to the Pleiades,
since their turn--off is located in the same position in the
color-magnitude diagram (Vidal 1973). Due to several properties, 
it  has  a promising future as a target for this type of
studies. Although it is moderately far away --(m-M)$_0$=9.7
magnitudes, compared with (m-M)$_0$=5.36 in the case of the Pleiades, 
Robichon et al. (1999)--, it has a clear 
advantage over other associations: its richness. It is one of 
the richest nearby young clusters, with a total mass
estimated between 1600 and 3200 M$_\odot$ 
(Leonard \& Merritt 1989).
Membership of those (few) stars hotter than 
about 5700 K can be checked against the excellent proper 
motion membership study of McNamara and Sekigushi (1985).
In addition, 
several optical  surveys have been recently conducted, 
producing long lists of candidate members (Sung \& Bessell 1999;
von Hippel 2000; Sarrazine et al. 2000).
The recent study by Sung \& Bessell (1999) yields 
(m-M)$_0$=9.60$\pm$0.10, E(B--V)=0.255$\pm$0.024 and an age
of 200$_{-100}^{+200}$ Myr, somewhat older than the turn--off 
age of the Pleiades, 80 Myr. Similar results have been derived
by Sarrazine et al. (2000), including
(m-M)=10.16$\pm$0.01,
E(B--V)=0.198$\pm$0.008,
and an age of 160$\pm$40 Myr.

 For some time now, we have been studying exhaustively
this cluster, tackling problems such as the mass function of the
cluster (Barrado y Navascu\'es et al. 1999, 2000a), rotational periods,
 and  multicolor photometric surveys
(Barrado y Navascu\'es et al. 2000a). Moreover, since the cluster
covers less than 1 sq.deg., it is a perfect candidate for multi--object
spectroscopy. We have taken advantage of the WIYN/HYDRA
multi--fiber spectrograph and observed a sample of M35 candidate
members, with the goal of identifying {\it bona fide} members,
 determining the pattern of lithium
depletion and rotation for them, and establishing  the
connection between these two last properties (rotation and
lithium).  We present the data in
Section 2, whereas the analysis and comparison with open
clusters, including the Pleiades, are discussed in 
Section 3. A summary and the main  conclusion can be found in 
Section 4.

%%%%%%%%%%%%%%%%%%%%%%%%%%%%%%%%%%%%%%%%%%%%%%%%%%%
%%%%%%%%%%%%%%%%%%%%%%%%%%%%%%%%%%%%%%%%%%%%%%%%%%%
%%%%%%%%%%%%%%%%%%%%%%%%%%%%%%%%%%%%%%%%%%%%%%%%%%%
\section{The data}

\subsection{M35 observations and data reduction}

We selected our initial sample of M35 photometric candidate
from Barrado y Navascu\'es et al. (2000a). In that paper, we 
produced a  list of cluster candidates  based on their 
location in a V,I$_c$  and I$_c$,R$_c$ color-magnitude diagrams.
 Our targets  have colors and magnitudes  in the ranges
V=14.5--17.5 and  (V--I)$_c$=0.85--1.55.
Figure 1 displays the location in a color-magnitude diagram of
our target stars. In total, we observed 76 candidate members of
the cluster. Table 1 lists the names and positions
of these stars, as well as the photometry (our data from 
Barrado y Navascu\'es et al 2000a and a compilation of data from
Sung \& Bessell 1999). The last column in Table 1 indicates 
the membership status of these candidates (see Section 3.1).

Our  candidate members were observed spectroscopically
using WIYN/HYDRA, a multifiber spectrograph, capable of
observing 97 targets simultaneously.
We took 6 individual exposures of $\sim$2 hours each, over 2 different
nights.

The reduction was carried out with the
IRAF\footnote{IRAF is distributed  by National 
Optical Astronomy Observatories, which is operated by
the Association of Universities for Research 
in Astronomy, Inc., under contract to the National
Science Foundation, USA} environment, using standard procedures
(bias subtraction, flat--fielding, extraction of the individual
spectra and wavelength calibration). IRAF contains a specific
package which performs optimally  all these functions in a
standardized way, denominated "dohydra".
The final free spectral range is 6440-6850 \AA, with  a
resolution of R$\sim$20\,000, as measured in the comparison lamp
spectra (2 pixels). 
Since we collected spectra in two consecutive nights, and within 
each night the observations were split in individual exposures
two hours long each, we processed each individual set of spectra
in  a complete independent way. Once the exposure was reduced and the
individual spectra were extracted for each star, we combined
all the spectra corresponding to the same star using a median
filter, in order to remove remaining cosmic rays. 
The signal--to--noise ratios range from 40 per pixel for our faintest
objects, to 160 in the case of  the brightest.
This is a remarkable feat for stars which are at (m-M)=10.4
magnitudes. Some examples of spectra of our sample of 
M35 candidates are shown in  Figure 2.
The effective temperature decreases upward. Note the
obvious change of the H$\alpha$ equivalent width and profile,
and the presence of the lithium doublet at  \ion{Li}{1}\,6708\AA.

\subsection{Data from other clusters.}

In the next sections we will analyze the data we have derived for 
M35 candidate members. We will also compare our data
 with data collected from a
number of open clusters, namely the Pleiades, NGC2516,
M34 --NGC1039--, NGC6475, Ursa Majoris moving group --UMaG--
 and the Hyades. The turn--off ages of these clusters lie 
in the ranges 70--100, 
120--150, 200, 220, 300 and 600-800 Myr, respectively.
At the present time, 
our group has estimated the ages of the open clusters 
IC2391 (Barrado y Navascu\'es, Stauffer \& Patten 1999),
Alpha Per (Stauffer et al. 1999), and  the Pleiades (Stauffer,
Schultz, \& Kirkpatrick 1998), based on the lithium depletion
boundary technique (LDB) in very low mass members of each
cluster. These LDB ages are $\sim$50 \% older than the turn--off
ages. However, they agree with turn--off ages computed with a
moderate amount of core overshooting (Ventura et al. 1998). In
fact, Stauffer \& Barrado y Navascu\'es (1999) have been
able to define a new age scale for young open clusters based on
the lithium dating technique, in opposition to the Upper Main
Sequence age scale, such as that from Mermilliod (1981).
Note that in the new LDB age scale, the Pleiades is 125$\pm$5
Myr. Although there are now very deep photometry of M35, reaching
the lithium depletion boundary of the cluster (Barrado y
Navascu\'es et al. 1999, 2000a), the distance modulus of the cluster
is so large that it poses an extraordinary challenge to detect
lithium at the bottom of its Main Sequence even with the largest 
telescopes and a significant amount of observing time.
 Regardless of the age scale we use, the conclusions of this
study remain the same, providing we use ages in the same scale
for each cluster. Therefore, we will work with the 
 turn--off age scale, keeping in mind that the real ages may
be a 50 \% older.

The data for these clusters we are going to compare come from
several sources. In the case of the Pleiades, photometry,
rotation, activity, and lithium equivalent widths come from
Soderblom et al. (1993a), Garc\'{\i}a L\'opez, Rebolo, \&
Mart\'{\i}n (1994), Jones et al. (1996) and Jeffries (1999).
Additional photometric data were selected from 
Johnson \& Mitchell (1958),
 Stauffer (1982, 1984),  and Stauffer \& Hartmann (1987).
 NGC2516 data come from Jeffries, James,   \& Thurston (1998). 
Jones et al. (1997) is the source of M34 data, whereas 
James \& Jeffries (1997) provides the information for NGC6475.
Data for UMaG were collected by Boesgaard et al. (1988) and
Soderblom et al. (1993c).
Finally, Hyades data were selected from a
number of sources, including
Duncan \& Jones (1983), Boesgaard \& Tripicco (1986),
Rebolo \& Beckman (1988),  Soderblom et al. (1990), 
Thorburn et al. (1993),  Barrado y Navascu\'es \& Stauffer
(1996). We have to point  out
that in some cases, these databases only provide the (B--V) 
or (V--I)$_k$ color
indices, whereas our M35 sample was observed in the V,I$_c$
filters. In general, we look for additional data in these last
two bands, losing some stars which do not have this type of
data. In some cases, we only could find data in the Kron
system, instead the Cousins system. Then, the (V--I)$_k$ color
was transformed into (V--I)$_c$ color using  Bessell \& Weis
(1987) transformation.
However, for other comparisons, we preferred to keep all the
original sample (such as in some comparisons between M35 and the
Pleiades). In this case, 
we  converted those (B--V) Pleiades  values into (V--I)$_c$
 colors using an average relation between these colors
which was computed using Pleiades stars observed in both colors.
This transformation introduces a bias, due to the inhomogeneous 
reddening for some Pleiades stars (Soderblom 1993b).
We emphasize that all the data for these seven clusters were treated in 
an homogeneous  way (for instance,
the derived lithium abundances).

%%%%%%%%%%%%%%%%%%%%%%%%%%%%%%%%%%%%%%%%%%%%%%%%%%%
%%%%%%%%%%%%%%%%%%%%%%%%%%%%%%%%%%%%%%%%%%%%%%%%%%%
%%%%%%%%%%%%%%%%%%%%%%%%%%%%%%%%%%%%%%%%%%%%%%%%%%%
\section{Analysis}

%%%%%%%%%%%%%%%%%%%%%%%%%%%%%%%%%%%%%%%%%%%%%%%%%%%
\subsection{Radial velocities and membership}

Our spectral resolution 
(R$\sim$20,000) allows us to compute radial velocities to an 
accuracy of $\sim$1 km/s. 
This was achieved by
measuring the positions of strong lines  (from iron and
calcium) on the spectra and deriving the shifts respect their
rest  wavelengths. We also performed cross-correlation
against  M35 candidate members. 
HYDRA is a bench spectrograph and, therefore,
the wavelength calibration  is very precise
(that is, it always yields the same value).
 In any case, even if
there is any shift having other origin than the relative velocity 
of the M35 candidates  and Earth, this unlikely shift should be the
same  for all stars in our sample, since  a particular 
exposure for a star was obtained simultaneous to the rest of 
candidates. Since we have several exposures which were taken
 during  two consecutive  nights, we have measured radial
velocities in each of them. Therefore, we have 11 data-points
for each M35 candidate member.

Based on the radial velocity information, we have classified
our M35 candidate members in three different categories:

\begin{itemize}

\item Probable members, single stars. These stars do not present
variability on the radial velocity (at least in the time scale
of our series of observations, two days) and the measured values are well
within the average radial velocity of the cluster 
($<$RV$>$$_{\rm M35}$=--8.0$\pm$1.5 km/s).
There are 39  objects in this group.
 Average radial velocities and dispersions are listed in column \#2 
of Table 2 for these stars.

\item Possible members, spectroscopic binaries. This group
includes all the candidates with variable radial velocity.
Some of them show lines arising from both components of the
system.
Since we have obtained only a handful of spectra per star, we are
not able to estimate the radial velocity of the baricenter
and, therefore, to determine their membership status. 
There are 13 objects in this category, including 3 SB2 binaries

\item Probable non-member. Those stars whose   radial velocity
is constant and different to the cluster average have been
classified as non-members based on this criterion. Moreover, the
majority of these stars  have a remarkable weak
 \ion{Li}{1}\,6708\AA{ } doublet. 
Most field stars are likely 
to be older than the very young M35, and are thus likely to 
have depleted their Li, consistent with our findings.

\end{itemize}

Note that we only have used lithium
as a membership criterion to confirm the classification as
 a non-member of stars falling in this group. Figure 1 depicts
all
the stars in our sample, probable members, possible SB members
and
non-members, as solid circles, empty circles and crosses,
respectively. Since our goal is to study the properties of dG--dK
members of the M35 cluster (rotational velocities, metallicity,
activity and lithium abundances), we do not consider further
the non-member. See Barrado y Navasc\'ues
 et al. (2000a) for more
information about them.
 We defer the discussion about the possible
binaries of M35 to a forthcoming paper, which will analyze the
lithium abundance and photometric properties of the components
in the same fashion as Barrado y Navascu\'es \& Stauffer (1996)
and Barrado y Navascu\'es et al. (1997b), where we presented an
exhaustive and comprehensive analysis of 
long period and tidally locked binaries belonging to the Hyades and M67
open clusters. 
For the special 
role that short period binaries play in testing models of 
lithium depletion caused by rotationally induced mixing, see 
discussion in Deliyannis (1990); 
for related discoveries of high-Li short binaries, see,
 a) in  the Hyades (Thorburn et al. 1993),   
 b) in M67  (Deliyannis et al. 1994), and
 c) in the field and metal-poor 
stars (Ryan and Deliyannis 1995).
For a general description from the phenomenological
point of view of the role of binarity on the
lithium depletion phenomenon, see Barrado y Navascu\'es (1998).

%%%%%%%%%%%%%%%%%%%%%%%%%%%%%%%%%%%%%%%%%%%%%%%%%%%
\subsection{Rotational velocities}

Rotational velocities were derived by cross-correlation
technique. First, we selected one spectrum with very narrow
lines, and spinned it up using rotational broadening functions
with rotational velocities from 10 km/s to 90 km/s, with
increments of 10 km/s.  These artificially broadened
spectra were cross--correlated with the candidate members, using
the spectral range 6600--6730 \AA.
The {\it v~sini} values  are listed in the last column of Table 2.

Figure 3a depicts our measured  rotational velocity against the 
dereddened (V--I)$_c$ color index for all the probable members of
 M35.
 Solid circles represent actual measurements, whereas solid
triangles correspond to upper limits. As a comparison, we have
included data from the Pleiades, which were selected from Soderblom
et al. (1993a). This sample appears as  crosses in the figure. 
 Despite of its youth, M35 lacks  rapid rotators in this color
 range. Only a handful of the warmer stars in the sample have
projected rotational velocities well above the detection limit.
Their velocity ranges  between  20 and 30 km/s.
 Only one star in the cooler end (M35-5376)
has a high rotational velocity. (In fact, the largest among
our sample of probable  members of the cluster.)
However, stars belonging to the Pleiades open cluster
show a very different pattern in their distribution of rotational
velocities. A significant amount of them have values larger than
50 km/s, even reaching 130--140 km/s in two cases. 
Since M35 and the Pleiades have, allegedly, the same age,
this situation defies, till certain extent, our understanding of
the age-rotation connection. It should be taken into account that
the Pleiades
(V--I)$_c$ values where derived from   (B--V) colors. 
To avoid any possible bias introduced by this process,  
we have re-plotted the data in Figure 3b using measured (V--I)$_k$
colors for  Pleiades stars and transforming them into  (V--I)$_c$.
 Some of these stars have disappeared
from scene, due to the lack of adequate photometry. More
important, some of these Pleiades stars have been shifted
red-wards, and they lie now on top of the M35 cool rapid rotator.
The lack of concordance between (B--V) and (V--I) photometry
has been noticed by King, Krishnamurthi \& Pinsonneault (2000),
who pointed out that it could be due to errors in one passband
or to increased red flux from stellar spots.
In fact, they found, in their words ``... strong evidence that our
Teff values ... are affected by activity level.''
But the conclusion still holds, the Pleiades has a significant
amount of fast rotators which are not present in M35.
Other parameters might be working, such as a difference in metallicity
(see next subsection) and/or a different initial angular momentum
distribution (it would have been more uniform in M35).
An additional piece of information can be extracted from the
comparison between M35 and NGC2516 data, shown  in Figure 3c.
The similitude of both distributions is remarkable, an
understandable phenomenon if, as pointed out by Sung \& Bessel
(1999), M35 is  200$_{-100}^{+200}$ Myr, and therefore closer 
in age to NGC2516 than to the Pleiades.

%%%%%%%%%%%%%%%%%%%%%%%%%%%%%%%%%%%%%%%%%%%%%%%%%%%
\subsection{The metallicity of M35}

The lithium  depletion  depends, among other parameters, on age, 
rotation and  metallicity (see recent reviews by Balachandran
1995; Pinsonneault 1997; Mart\'{\i}n 2000). 
Therefore, in order
to interpret correctly our observations, we have to take into
account the metal content, specially if we intend to infer 
conclusions by comparison with other clusters.
To  our knowledge, there is no previous spectroscopic study of
the 
metallicity of this cluster. Sung \& Bessell (1999), based on
photometric data (UBVI filters), estimated [Fe/H]$_{\rm M35}$=-0.3.
Gratton (2000), using DDO photometry, estimated [Fe/H]$_{\rm M35}$=-0.28.
We have obtained an independent estimate of the M35 metallicity 
based on spectral synthesis. We have used
the MOOG spectral synthesis program (Sneden 1973) and 
 Kurucz (1992) model atmospheres. 
 In all cases, we used a gravity of 
log~g=4.5. The gf-values were derived by Balachandran
(private communication), via spectral synthesis of 
solar equivalent widths (Kurucz  et al. 1984).
We applied these data to a 
 subsample of our {\it bona fide}
 M35 single probable members.
 This group includes the 9 brightest-warmest stars,
with the highest S/N. 
The spectrum of some of these stars can be found in Figure 2.

We measured the equivalent widths of several
 iron lines present in the observed spectral range (about 10
per star, see Table 3).
We acknowledge that the number of lines is small, 
due to the short free spectral range, and all of them   belong to the
ion \ion{Fe}{1}. These facts introduce uncertainties in the final estimate. 
The [Fe/H] value was derived in two different ways: (i) We used
the effective temperature evaluated from the photometry
(temperature scale from Bessel 1979, with values derived from (V--I)$_c$,
listed in Table 1).
A model atmosphere with this temperature was introduced in MOOG
as input in order  to 
derive the iron abundance for each line. We list the average
value in columns \#2 and \#3 of Table 4. 
(ii) We adjusted the effective temperature (spectroscopic
temperatures) with MOOG
in order to obtain the  similar values of [Fe/H] for all lines at
 different excitation potentials. The results, as well as their
standard deviations, are listed in columns \#4 and \#5 of Table 4.

Figure 4 shows both sets of  results  (data derived using 
spectroscopic  and photometric temperatures appear as solid and
empty circles). The average values in both cases are very similar,
within 
the uncertainties. We have adopted a metallicity for the
M35 cluster
of [Fe/H]$_{\rm M35}$=--0.21$\pm$0.10 dex, corresponding to
the average value of our individual metallicities.
 In order to compare with
the Pleiades 
open cluster, we  proceeded in the same way with several Pleiades
members  of the same color  range
(spectra kindly provided by D. Soderblom).
In this case, the result is 
[Fe/H]$_{\rm Pleiades}$=+0.01 dex. This value can be compared
with those results for the Pleiades by
Cayrel, Cayrel de Strobel, \& Campbel (1988), +0.13 dex;
Boesgaard \& Friel   (1990), --0.034$\pm$0.024 dex;
Gratton (2000), --0.03$\pm$0.06 dex;
and  King et al. (2000),  +0.06$\pm$0.06 dex. 
Figure 4 also shows our two average  metallicities for
M35 and the Pleiades (solid and dashed lines, 
respectively).  We conclude that M35 is slightly metal--poor
compared with  the  Pleiades.

%%%%%%%%%%%%%%%%%%%%%%%%%%%%%%%%%%%%%%%%%%%%%%%%%%%
\subsection{The Age of M35}

Vidal (1973) estimated, based on the turn--off of the Pleiades
and M35 massive stars, that both clusters have the same age.
However, this comparison was made using photographic data.
Deep, accurate CCD photometry for M35 has not been available 
until very recently (Barrado y Navascu\'es et al. 1999;
Sung \& Bessel 1999; 
von Hippel at al.  2000; Barrado y Navascu\'es et al. 2000a).
This wealth of new data allows an accurate age determination.
Sung \& Bessell (1999), using  isochrone fitting in multiple
color-magnitude diagrams, have derived an age of
200$_{-100}^{+200}$ Myr,
significantly older than the Pleiades standard age of 80 Myr.
The isochrone fittings of the  red giant probable member and
the two red giants possible members give 125 and 200 Myr,
respectively.
The same qualitative results have been obtained by von Hippel et
al. (2000) fitting the MS of the cluster.
Moreover, Sarrazine et al. (2000) have estimated, using
several color-magnitude and color-color diagrams in the UBVRI
bands, that the age is 160$\pm$40 Myr. As comparison, they applied the
same technique to the Pleiades and derived a turn--off age of
80$\pm$20 Myr.
 Mysore (1999) analyzed a sequence of M35 V band images, identified a set
 of probable M35 members, and searched for photometric variability due to
 starspots among the G and K dwarf cluster members.  Despite being sensitive
 to variability sufficient to easily identify stars like the rapidly rotating
 K dwarfs in the Pleiades (which have typical light curve amplitudes in
 V of about 0.1 mag), she found no certain M35 variables.  She concluded
 from this that M35 is likely to be significantly older than the Pleiades.

These age ranges fully agree with our own findings concerning
  the distribution of rotational velocities and,
as we will see, the lithium depletion.
 Therefore, we will assume that M35 is
a 175 Myr old cluster, the average age of the 
three M35 red giants.

%%%%%%%%%%%%%%%%%%%%%%%%%%%%%%%%%%%%%%%%%%%%%%%%%%%
\subsection{Lithium abundances}

\subsubsection{The lithium equivalent widths}

We have  measured  
 the lithium \ion{Li}{1}\,6707.8\AA{ } equivalent widths 
--W(\ion{Li}{1})-- for all our M35 candidates.
 Due to our spectral resolution, 
the high lithium abundances and  the  rotation velocities,
 the lithium feature is blended with 
\ion{Fe}{1}\,6707.4\AA. Therefore, we removed this small
contribution using an empirical relation between color and
 W(\ion{Fe}{1}) derived by Soderblom et al. (1993a).
We list the blended and deblended equivalent widths 
W(Li$+$Fe) and W(Li) in columns \#3 and \#4 of Table 2.
Before analyzing  the pattern of lithium depletion in M35,
let's examine first how the lithium equivalent widths depend on
the color, dependence depicted in Figure 5.  Solid circles and
crosses represent M35 probable members and  Pleiades data from
Soderblom et al. (1993a), respectively. The  (V--I)$_c$ color
indices used in
panel a were derived from (B--V), whereas
those shown in panel b come from (V--I)$_k$ values (see section
2.2). It is easy to appreciate the scatter present in the
Pleiades data. For a discussion about the origin of this 
spread, see Soderblom et al. (1993a), 
King, Krishnamurthi \& Pinsonneault (2000),
 Barrado y Navascu\'es et al.  (2000b),    
and Section 3.5.3.
This scatter is more conspicuous for stars
redder than (V--I)$_{c,0}$=0.8. On the other hand, M35 stars 
in the range 0.6$<$(V--I)$_{c,0}$$<$1.0
 have, within the errors, a tighter relation between 
W(\ion{Li}{1}) and the color index.
 However, as it happens in Pleiades stars, there is an important scatter
for cooler M35 stars. Moreover, on average, M35 stars show a
smaller lithium equivalent widths for a given color, when 
compared with the Pleiades.
 If both clusters have the same age, this situation is difficult
to understand (this also was the case of rotational velocities,
with Pleiades stars rotating faster than equivalent M35 stars).
Why do M35 stars have smaller W(\ion{Li}{1}) than their Pleiades
counterparts? Why is the relation  better defined in M35? What
is the origin of the scatter around (V--I)$_{c,0}$$\sim$1.2?
In the next subsections we will try to shed some light over these
problems.

\subsubsection{The Teff-A(Li) plane}

We have derived lithium abundances for all stars discussed here
using the curves of growth computed by  Soderblom et al. (1993a). 
After measuring the lithium equivalent widths (or gathering them
from the literature in the case of the 
other open clusters, see Section 2.2), the following step was used to 
estimate effective temperatures. Since temperature scales based
on (B--V) colors are more sensitive to metallicity than those  
based on (V--I), and our clusters have different metal content
([Fe/H]$_{\rm M35}$=--0.21,
[Fe/H]$_{\rm Pleiades}$=$+$0.01
 --both from this paper, see section 3.4--,
[Fe/H]$_{\rm NGC2516}$=--0.32 --Jeffries et al. (1997)--,
    [Fe/H]$_{\rm M34}$=--0.29 --Gratton 2000--, 
[Fe/H]$_{\rm NGC6475}$=+0.11 --James \& Jeffries 1997--,
[Fe/H]$_{\rm UMaG}$=--0.079 or --0.085 --Boesgaard et al. 1988
and Boesgaard \& Friel 1990, respectively-- ,
[Fe/H]$_{\rm Hyades}$=$+$0.16, or $+$0.127, or $+$0.13  
--Cayrel, Cayrel de Strobel, Campbel 1988; Boesgaard \& Friel 1990;
and Gratton 2000,
respectively--),
 we derived effective temperatures from (V--I)$_c$ colors, after
Bessell (1979) temperature scale. The effective temperature 
values are listed in columns \#6 and \#8 of Table 2.
In the first case, the temperature was computed using our 
values of (V--I)$_c$ (Barrado y Navascu\'es et al. 2000a), 
whereas in the second case we used the color indices 
selected from Sung \& Bessell (1999).
These  two sets  of effective temperatures and the lithium equivalent 
widths were used to derived the lithium abundances listed in
columns \#7 and \#9 of the same table.
As can be seen, there is not relevant differences in both cases.
(i.e., the scatter introduced on the final lithium abundances 
by differences in the photometry of individual M35 stars  is
quite small.)
From now on, we will used the effective temperatures and
lithium abundances derived with our photometry. 
Errors in the computed abundances were estimated taking into
account the uncertainties in effective temperatures and lithium
equivalent widths.

Figures 6a and 6b are like Figures 5a and 5b, but in this case
we depict the abundance --in the customary scale where
A(Li)=12+log(Li/H)-- against the effective temperature.
Note that the temperatures corresponding to Pleiades stars
which are shown in Figure 6a come directly from Table 1 of
Soderblom et al. (1993a) and, therefore, they were computed 
from (B--V) colors. There are three features in these figures
which are worth discussing:

1.- Independently of the origin of the Pleiades temperatures, 
it seems clear that, for a given spectral type, M35 stars
have depleted more lithium than similar Pleiades stars. This
phenomenon is more conspicuous in the range 6000-5500 K, where
the Pleiades show abundances close to the cosmic value
--A(Li)$_{\bf cosmic}$=3.2, Rebolo (1989), Anders \& Grevesse (1989);
Mart\'{\i}n et al. (1994)--,
 whereas M35 members have abundances below A(Li)=3.0 dex. If indeed
the Pleiades and M35 have the same age, this situation 
arises some concerns on our understanding of the processes taking
place during the lithium depletion in pre-MS stars and its relation
with age. The distribution of rotational velocities is similar
for both clusters in this temperature range, and a rotational
effect cannot be responsible of the differences in lithium
abundances. The difference in metallicity cannot also be called
upon as the main actor,  since the difference is small and
it would act in the opposite direction: the bottom of the
convective envelope would be closer to the surface in the case
of M35, making more difficult the lithium depletion.

2.- Pleiades stars in the range 5300-5000 K have a very relevant
lithium spread, related to rotation and activity 
(Butler et al. 1987;
Soderblom et al. 1993a; Garc\'{\i}a L\'opez, Rebolo, \&
Mart\'{\i}n 1994), larger than 1 dex.  
This  scatter  does not appear in the case of  M35 members.
However, this situation does not prove the lack of connection
between lithium depletion and rotation/activity, since these M35
members are very slow rotators. That is, there is no slow rotator
with high abundance and {\it  vice versa}, a fact which either
would contradict
the  relation between lithium depletion and rotation,
 or would indicate that the rotational history is playing an important 
role (in opposition to a sharp connection rotation--lithium
for a given age).
See section 3.5.3 for an additional discussion about the lithium
spread.

3.- Stars at lower temperature (Teff$\sim$4500 K) do show a 
large spread ($\Delta$Li$>$1.0 dex). 
The spread is quite apparent if the spectra of these stars
are compared (see Figure 7a,b).
This spread is marginally
related to rotation, but unfortunately our sample only contains 
one cool fast rotator and it is safer not to derive any conclusion
at the present stage.
Only additional data can establish the origin  of this spread and 
its relation with rotation and activity.

There are four M35 probable members with only upper limits to their
lithium abundances. Three of them are quiet cool, and their low abundances,
below A(Li)=0 dex, can be explained by the dependence of the lithium
depletion on mass, as happens in slightly cooler Pleiades stars. 
However, one among these stars, \#5190, is warmer and its behavior
 clearly departs
from that characteristic of similar M35 members at the same temperature.
 This 
is probably an indication of non-membership. However, since we only used 
radial velocities as membership criterion, we have left this star in 
our sample of members of the M35 cluster.

\subsubsection{The connection lithium-rotation-activity}

During the last decade, much has been discussed about the 
nature of the lithium abundance spread in Pleiades early K dwarfs.
From the observational point of view, there are several pieces of evidence
which support the fact that  the derived 
abundances  are real, the spread is real, 
 and it is  connected with the stellar rotation
rate. Butler et al. (1987) were the first to 
notice the spread and it relation with rotation,
 followed by   Garc\'{\i}a L\'opez et al. (1991a,b).
Soderblom et al. (1993a) carried out a systematic analysis of
the lithium spread and its relation with rotation and activity, 
using different indicators such as H$\alpha$ and the \ion{Ca}{2}
 infrared triplet. They also considered the possible effect
of stellar spots 
on the apparent lithium equivalent widths and the derived abundances,
concluding that the spread was real, and that for a given color, those stars
rotating faster had the largest lithium abundances and chromospheric activity.
Jones et  al. (1996) provided additional data, including cooler
K spectral type stars. They confirmed the spread 
for early K and the break down of the relation for the fainter objects,
noticed by Garc\'{\i}a L\'opez et al. (1994).

However, other studies, either from the observational or
the theoretical point of views, have challenged these 
results and their interpretation.
Houdebine \& Doyle (1995)  computed a grid of model atmospheres
for M dwarfs in non local thermodynamical equilibrium, showing that
very high chromospheric activity  levels could 
modify the  strength of \ion{Li}{1}\,6708\AA.
Russell (1996) found that derived abundances for several Pleiades
stars using   \ion{Li}{1}\,6103\AA{ } 
 are different than those computed with 
\ion{Li}{1}\,6708\AA. The lithium spread disappeared with 
the former abundances. He interpreted the situation
as an indication of the effect of chromospheric activity 
on the \ion{Li}{1} line, in agreement with the results of 
Houdebine \& Doyle (1995), although Mart\'{\i}n (1997)
has pointed out the difficulty  of deriving abundances
using \ion{Li}{1}\,6103\AA{ } due to the presence of 
much stronger nearby lines and 
Stuik et al. (1997) affirmed that \ion{Li}{1}\,6708\AA{ }
and \ion{K}{1}\,7699\AA{ } are not very sensitive to the
presence of a chromosphere. They argued that the potassium feature
can be used as a proxy to understand the formation of 
the lithium line without the effect of the abundance spread.
Note that the \ion{K}{1}\,7699\AA{ } equivalent width does not depend on age,
since potassium is not depleted in the stellar interior, 
in opposition to what happens to lithium.
However, they could not establish if the  magnetic activity
can affect the \ion{K}{1}\,7699\AA{ } in Pleiades stars.
In the same line, 
Jeffries (1999), by monitoring several Pleiades stars,
 concluded that it is still
 unsafe to attribute the spread 
to real abundance differences, and King, Krishnamurthi \&
Pinsonneault (2000) presented some  additional evidences
 that the Pleiades  lithium dispersion 
is partially  due to stellar atmosphere effect.
Moreover, they found a strong correlation between 
stellar activity and lithium abundances (see also 
Fernandez-Figueroa et al. 1993), but the relation between rotation 
and lithium abundance  is not  one-to-one, making unlikely that
the reason of the lithium spread is rotation {\it per se}.
Our own simulations
(Barrado y Navascu\'es et al. 2000b)
indicate that stellar inhomogeneities (spots and active regions) 
could, indeed, introduce part of the lithium scatter, due to the 
combined  effect of the inhomogeneities  on the observed 
color (a star would appear to be cooler and less massive that 
it  really is) and the equivalent widths (the observed
lithium equivalent widths) would be increased in a significant amount, an 
effect noticed by Giampapa (1984).
 As shown by Barrado y Navascu\'es et al. (2000b),
 the presence on the surface of stellar
inhomogeneities can change in a significant way the observed 
colors and  lithium equivalent widths,  if the
activity rates are very high (and, with them, the filling factor
or stellar spots and active regions). Doppler imaging of active
Pleiades stars shows that they are, indeed, covered with huge
polar spots (Stout-Batalha \& Vogt 1999) of presumably
long life span.    The observed relation between lithium and
rotation could be due, to some extent, to the link between
rotation and activity. This situation does not
mean that there are not genuine differences in the lithium
abundances for stars of the same age and masses,
 and that these differences are related to the
rotation and the rotational history of the star, as proposed by
Deliyannis (1990), Deliyannis et al. (1990) and 
Pinsonneault, Deliyannis \& Demarque (1991, 1992a,b).
In fact, Barrado y Navascu\'es et al. (2000b) have shown that under
realistic conditions (when the filling factor of spots is less than
0.30), only part of the lithium spread can be attributed to 
surface inhomogeneities. 
Moreover,  it has been shown that tidally locked binary systems
(TLBS) in the Hyades and M67 have larger abundances than similar 
non-TLBS or single stars (Thorburn et al. 1993;
Deliyannis et al. 1994; Ryan \& Deliyannis 1995;
Barrado y Navascu\'es \& Stauffer 1986;
Barrado y Navascu\'es et al. 1997b). Since most of the rotational
periods of these Hyades and M67 TLBS are larger than any Pleiades
star,  the activity and the spot filling factor can not be
invoked to explain these differences in abundances, which seem to
be real. The same situation holds for chromospherically active
binaries of different ages and evolutionary stages
(Fernandez-Figueroa et al. 1993; Barrado y Navascu\'es 
1996, 1997; 
Barrado y Navascu\'es et al. 1994, 1997a,  and 1998).
Therefore, the real lithium  abundance differences due to
rotation {\it per se}  would  be smaller than
the values normally accepted.
 In this scenario, Pleiades rapid
rotators would lose most of their angular momentum before they
reach the M35 age, reducing their stellar activity. Therefore, 
the filling factor of stellar inhomogeneities would be reduced
considerably, and the observed photometry and equivalent width
of the star would be modify accordingly (the star would be moved
blue-ward in the color magnitude diagram and Teff--lithium plane,
due to the lack of cool spots). At this point, most of the 
lithium scatter would have disappeared. The remaining spread
would be due to the real effect of rotation on the lithium
depletion.

Figure  6 shows clearly that M35, an open cluster
significantly older than the Pleiades (Section 3.4), does not 
have a large scatter for early K stars. If the observed Pleiades
scatter is due to real abundance differences and the Pleiades
cluster represents the situation of M35 when it was  70-100 Myr old,
 it should be  explained how the lithium spread has disappeared
in these tens of million years. That is, all Pleiades stars for a given 
mass will  have to deplete their lithium in a specific amount
(more for those with the larger abundance) so they will end up
with the same abundance  once they reach M35 age (175 Myr). 
There are two possible scenarios which can account for this
situation: 
(i) If we assume that the  differences
in the lithium abundances for a given temperatures are
no real, that they are an artifact of the chromospheric
activity and the presence of inhomogeneities
on the stellar surface, as explained above, then 
 we would be comparing stars
with the same observed effective temperature, but different
masses. The most active star, being also the one with
the larger apparent abundance, would be, in fact, more massive.
Moreover, the observed lithium equivalent width would
not correspond to real one, due to the change  of the apparent
lithium equivalent width. Unfortunately, no model has been able to 
quantify properly these effects, in order to remove them
and derive real abundances for the Pleiades stars.
(ii) An alternative would be the combination of metallicity and 
the distribution of rotational velocities. As we have shown, 
M35 has a distribution of rotational velocities much 
tighter than the Pleiades. It metallicity is also smaller.
If we assume that the distribution of initial angular momentum
is tighter in metal--poor stars, 
the Yale models (Pinsonneault et al. 1990, 1992) would predict  
a tight dependency of lithium on temperature for M35, compare
with the Pleiades.

In any case, the high quality M35 data presented here, together with the 
rotational velocities and the tight relation between lithium abundance
and temperature, makes this young open cluster an ideal laboratory 
to test theoretical models proposed to explain the lithium depletion 
in late spectral type stars.

\subsubsection{Lithium depletion and  age}

Figures 8a-f contain several comparisons between M35
probable members and stars belonging to open clusters of
different ages. Namely, 
the Pleiades, NGC2516, M34, NGC6475, UMaG and 
the Hyades; with ages 80, 150, 200, 220, 300 and 800 Myr,
 respectively. Note that other age values can be found in the
literature. 
For the origin of the data and the way they were 
homogenized, see section 2.2. 
Temperatures and lithium abundances were derived in the
same fashion for all databases.
As pointed out by a number of studies (e.g. Balachandran 1995,
Pinsonneault 1997; Mart\'{\i}n 2000; Jeffries 2000; Deliyannis 2000,
 and references therein),
 open clusters of different ages show a
characteristic pattern of lithium depletion: the older the age,
the smaller the abundance for a given mass. In the particular
case of these seven clusters, it is clear that Hyades stars have
depleted their  lithium in a much larger amount than M35, whereas
Pleiades stars (specifically, those of early G spectral type) have
kept most of their original content, when M35 stars have already
destroy part of it. NGC2516 data is more sparse, but the data
seem to indicate that the distribution of lithium abundances of
NGC2516 and M35 stars are alike. Note, however, that NGC2516 shows a 
spread in the abundances in the Teff range 6000-5000 K not present in
the case of M35.  Figure 8c seems to indicate that M34 members
could have 
depleted more lithium, on average, than M35 counterpart. 
However, the lithium spread is quiet large in the case of M34 stars.
Since the metallicities of both cluster quite similar, 
this situation can  be 
interpreted  as a sign that M35 is slightly younger than
M34 (i.e., younger than 200 Myr).
The comparison of the distribution of lithium abundances with clusters
in this age range, such as NGC6475, yields similar results: abundances
are distributed in similar way, but in this last case, as happens
in M35, there is no spread, the relation lithium abundance and effective 
temperature is quite tight. Again, average abundances of NGC6475 are lower 
than those characteristic of M35, for a given age.
Abundances of UMaG  members are more  depleted compared with 
M35 and Hyades members have
 abundances at least 1.5 dex lower than M35 members.

If we accept that, indeed, M35 is significantly older than
the Pleiades, these results fit in the picture, and it
 is reasonable to
assume that the main parameter involved in the lithium depletion
is still  age (for a given stellar mass). Metallicity, at least
in this range of values, is a second order effect. 
In fact, rotation  {\it per se} could play a secondary role also
in the lithium depletion phenomenon. It is remarkable 
 the fact that between 80 Myr (the Pleiades
age) and 175 Myr (our adopted age for the cluster, see section
3.4), the differences between 
rotation rates for a given mass have disappear and, with them, 
the lithium abundances have come to the same value (assuming that
the Pleiades represents the M35 stage at 80 Myr). We interpret
the phenomenon from another perspective. We believe that
the Pleiades, in truth, represents the characteristic properties
of stars of its age, there is nothing abnormal about the cluster
to lead us to think otherwise. However, the differences in
abundances for a given mass, the scatter related with
rotation/activity might be due to the activity {\it per se}, at least
partially (see the discussion  in Section 3.5.3). 
Therefore, the pattern of lithium depletion  in M35 seems to confirm 
the assumed age of the cluster ($\sim$175 Myr) and the effect of rotation
 and stellar activity  on the apparent lithium abundance.

%%%%%%%%%%%%%%%%%%%%%%%%%%%%%%%%%%%%%%%%%%%%%%%%%%%
%%%%%%%%%%%%%%%%%%%%%%%%%%%%%%%%%%%%%%%%%%%%%%%%%%%
%%%%%%%%%%%%%%%%%%%%%%%%%%%%%%%%%%%%%%%%%%%%%%%%%%%

\section{Summary and conclusions}

We have collected multi-fiber, high resolution  spectroscopy of 76
 candidate members of the young cluster M35. Our spectra have
moderate to high signal--to--noise rations (S/N$\sim$40--160), a remarkable 
feat for stars located at (m-M)=10.4.
We have measured radial velocities and using these values as
membership criterion, we have cataloged 39 stars out of the original 76
as probable members of the cluster.
Another 13 stars are spectroscopic binaries and part of them could 
be members of the cluster.
 We also have measured rotational velocities
via cross--correlation. The comparison between the the rotation pattern
of members of M35 and the Pleiades indicates that,
 indeed, M35 is somewhat older, 
about 175 Myr, since it lacks fast rotators.  
Our high quality M35 spectra have been used to derive the metallicity
of the brighter and warmer members of the cluster. The result 
yields  [Fe/H]$_{\rm M35}$=--0.21$\pm$0.10, lower than
 the Pleiades metal content, 
[Fe/H]=+0.01.
Subsequently, we have derived  lithium abundances of the M35 members, 
and compared these values with members of other nearby clusters 
of different ages
and metallicities. We conclude that  at the age of M35, most of the lithium
spread observed in Pleiades stars is not longer present.
 We have interpreted this 
fact, at least partially,  as an effect of the photospheric spots
 on the apparent lithium abundance of Pleiades stars. Stellar rotation 
could also play some role. The data presented in this study are
perfectly suited to be used as a validity check of theoretical models involving
the lithium depletion.

%%%%%%%%%%%%%%%%%%%%%%%%%%%%%%%%%%%%%%%%%%%%%%%%%%%
%%%%%%%%%%%%%%%%%%%%%%%%%%%%%%%%%%%%%%%%%%%%%%%%%%%
%%%%%%%%%%%%%%%%%%%%%%%%%%%%%%%%%%%%%%%%%%%%%%%%%%%
\acknowledgements

DBN thanks the  {\it ``Instituto de  Astrof\'{\i}sica de Canarias''}
and {\it ``Ministerio de Educaci\'on y Cultura''}
 (Spain), and the {\it ``Deutsche Forschungsgemeinschaft''}
(Germany) for   their fellowships. 
CPD  gratefully acknowledges support for this work from the 
National Science foundation under Grant AST-9812735.
JRS acknowledges support from NASA Grant NAGW-2698 and 3690.
 This work has been partially suported by 
Spanish {\it ``Plan Nacional del Espacio''}, under grant ESP98--1339-CO2.
We appreciate the very useful comments and suggestions by the referee,
R.J. Garc\'{\i}a L\'opez.

\newpage

%%%%%%%%%%%%%%%%%%%%%%%%%%%%%%%%%%%%%%%%%%%%%%%%%%%
%%%%%%%%%%%%%%%%%%%%%%%%%%%%%%%%%%%%%%%%%%%%%%%%%%%
%%%%%%%%%%%%%%%%%%%%%%%%%%%%%%%%%%%%%%%%%%%%%%%%%%%

\newpage  %%%% Here come  the Figure captions %%%%%

\begin{center}
{\sc Figure Captions}
\end{center}

\figcaption{Color-magnitude diagram for M35 candidate members.
Note that the reddening --E(V--I)$_c$=0.21-- has not been
removed.
\label{fig1}}

\figcaption{Spectra of several M35 candidate members observed
with WIYN/HYDRA. 
The top spectrum corresponds to the faintest M35 member in our 
sample, whereas the bottom spectrum is the brightest. 
  The lithium feature at  6708 
\AA{ } is easily distinguished.
\label{fig2}}

\figcaption{Projected rotational velocity against unreddened
(V--I)$_c$ color.
M35 probable members as shown as filled symbols (circles for real
measurements
and triangles for upper limits). 
{\bf a} Comparison with Pleiades stars  (crosses). 
Values  for Pleiades stars derived from (B--V), see text.
{\bf b} Comparison with Pleiades stars  (crosses).
Values  for Pleiades stars derived from (V--I)$_k$, see text.
{\bf c} Comparison with NGC2516 stars  (crosses).
\label{fig3}}

\figcaption{Metallicity of the warmest M35 probable members
belonging to our sample.
Solid and open symbols represent temperatures derived via 
spectral synthesis and from the photometry, respectively. 
 The solid line represent the average M35 metallicity, whereas
the long-dashed line corresponds to the Pleiades.
\label{fig4}}

\figcaption{Comparison between the  \ion{Li}{1}\,6708\AA{ }
equivalent widths
of M35 probable members (solid circles) and Pleiades members
(crosses, data from Soderblom et al. 1993a)
. In panel  {\bf a}, Pleiades (V--I)$_c$ data was derived from (B--V),
whereas in panel  {\bf b} we represent Pleiades (V--I)$_c$  values 
computed from (V--I)$_k$.
\label{fig5}}

\figcaption{Lithium abundances against effective temperatures.
M35 probable members are shown as solid circles, whereas Pleiades
stars appear as crosses.
{\bf a} Soderblom et al. 1993a original data.
{\bf b} temperatures and abundances derived from (V--I).
\label{fig6}}

\figcaption{Spectra of our coldest targets (4400$>$Teff$>$4200).
 Note the effect of rotation and the different strength of
lithium feature.
\label{fig7}}

\figcaption{Lithium abundances against effective temperatures.
M35 probable members are shown as solid circles ($\sim$175 Myr).
Panels represent several comparisons with data belonging to 
several open clusters: {\bf a} The  Pleiades (70--100 Myr).
{\bf b} NGC2516 (120--150 Myr).
{\bf c} M34 (200 Myr).
{\bf d} NGC6475 (220 Myr).
{\bf e} UMa Group (300 Myr).
{\bf f} The Hyades (600-800 Myr).
\label{fig8}}

%%%%%%%%%%%%%%%%%%%%%%%%%%%%%%%%
\setcounter{figure}{0}
\begin{figure*}
\vspace{18cm}
\includegraphics{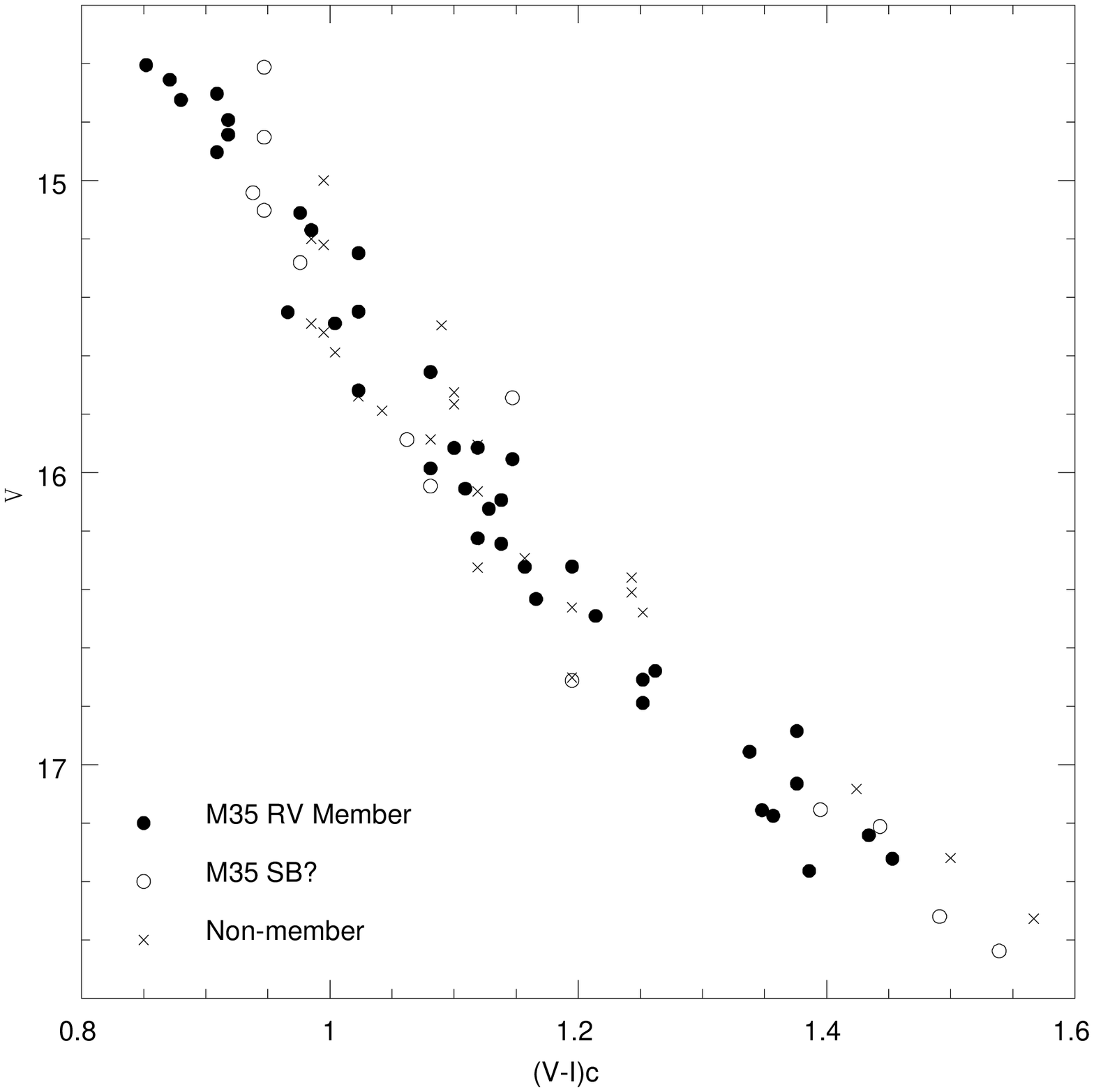}
\caption{}
\end{figure*}
%%%%%%%%%%%%%%%%%%%%%%%%%%%%%%%%

%%%%%%%%%%%%%%%%%%%%%%%%%%%%%%%%
\setcounter{figure}{1}
\begin{figure*}
\vspace{18cm}
\includegraphics{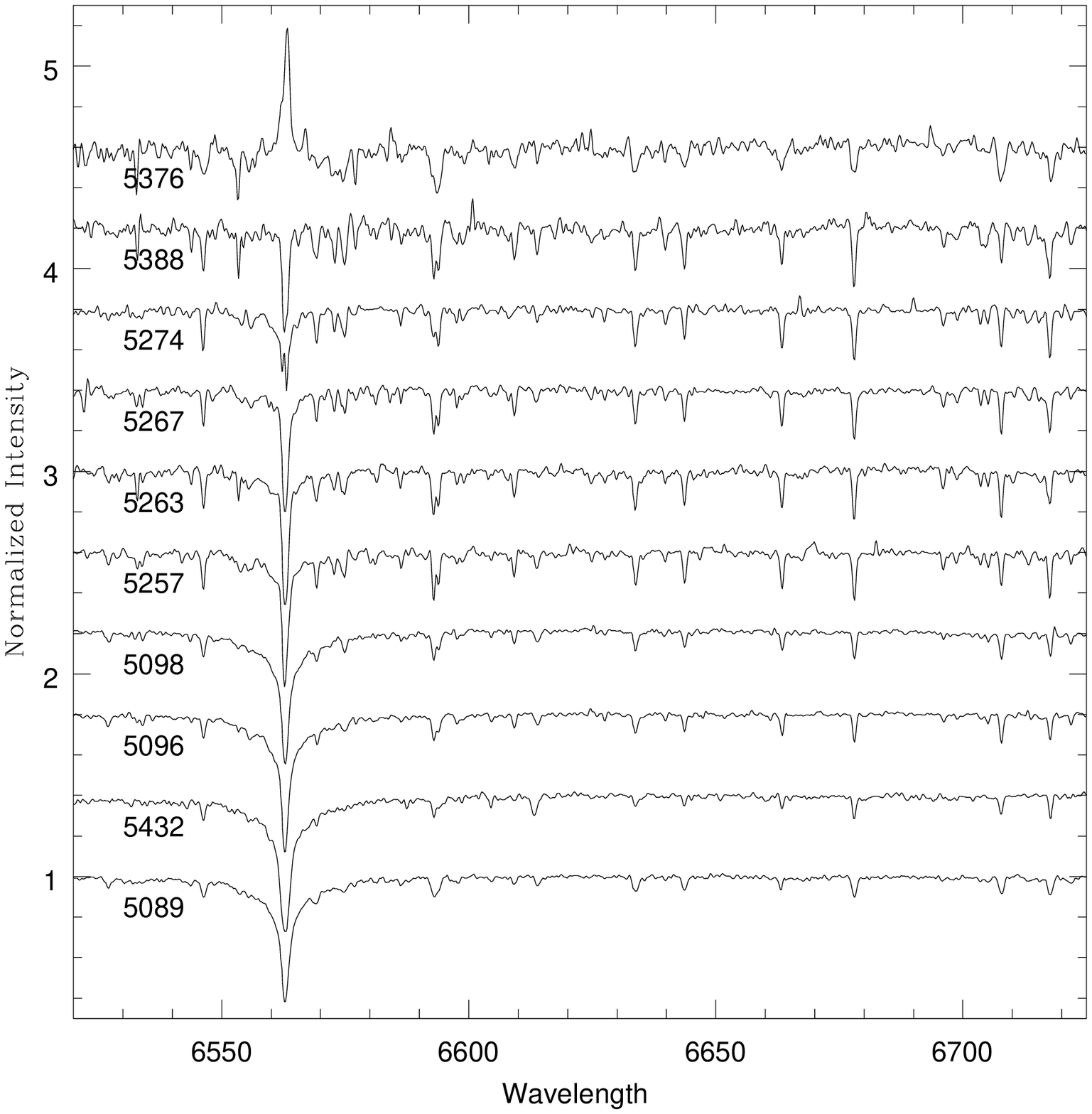}
\caption{}
\end{figure*}
%%%%%%%%%%%%%%%%%%%%%%%%%%%%%%%%

%%%%%%%%%%%%%%%%%%%%%%%%%%%%%%%%
\setcounter{figure}{2}
\begin{figure*}
\vspace{18cm}
\includegraphics{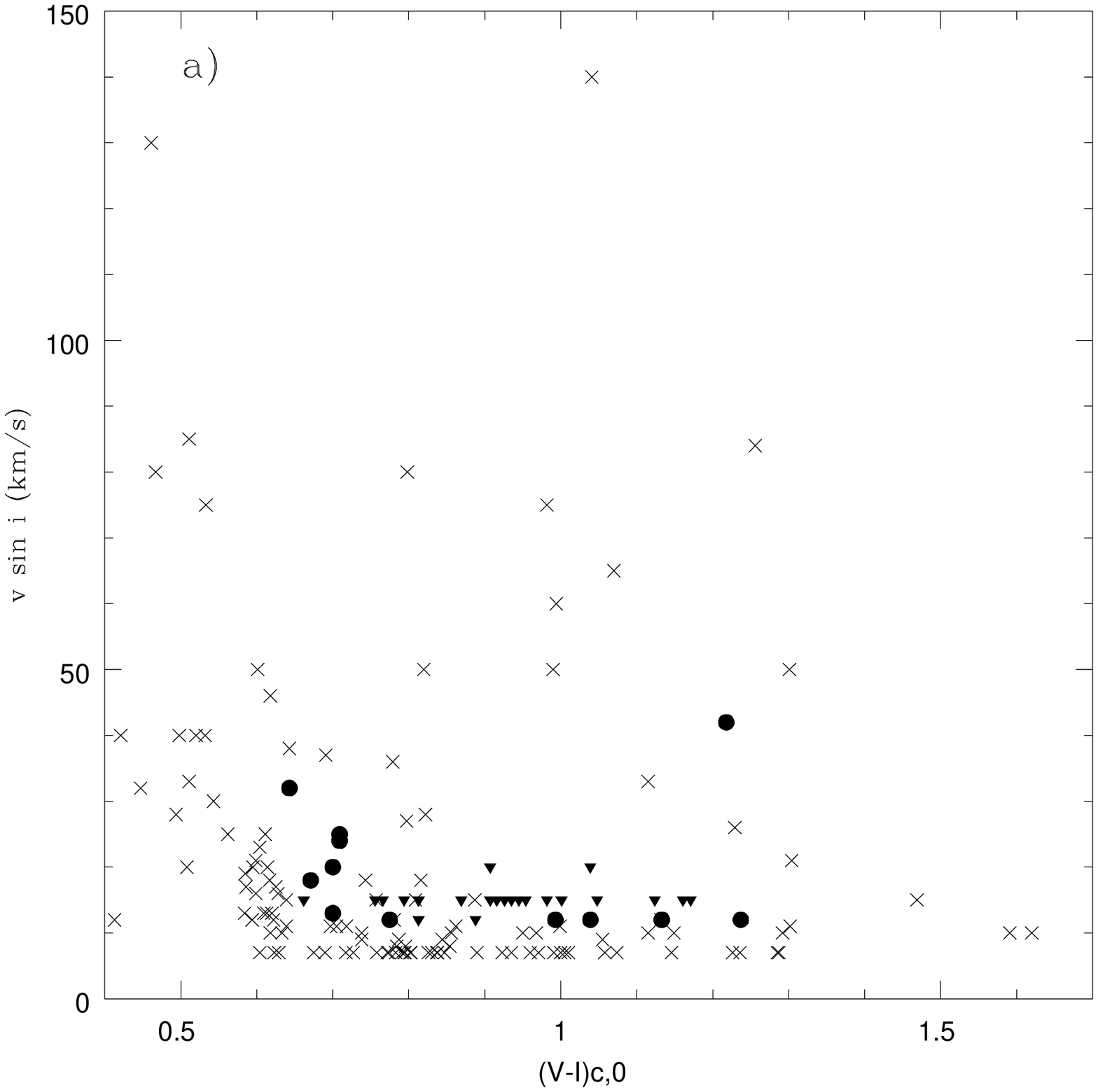}
\caption{{\bf a} }
\end{figure*}
%%%%%%%%%%%%%%%%%%%%%%%%%%%%%%%%

%%%%%%%%%%%%%%%%%%%%%%%%%%%%%%%%
\setcounter{figure}{2}
\begin{figure*}
\vspace{18cm}
\includegraphics{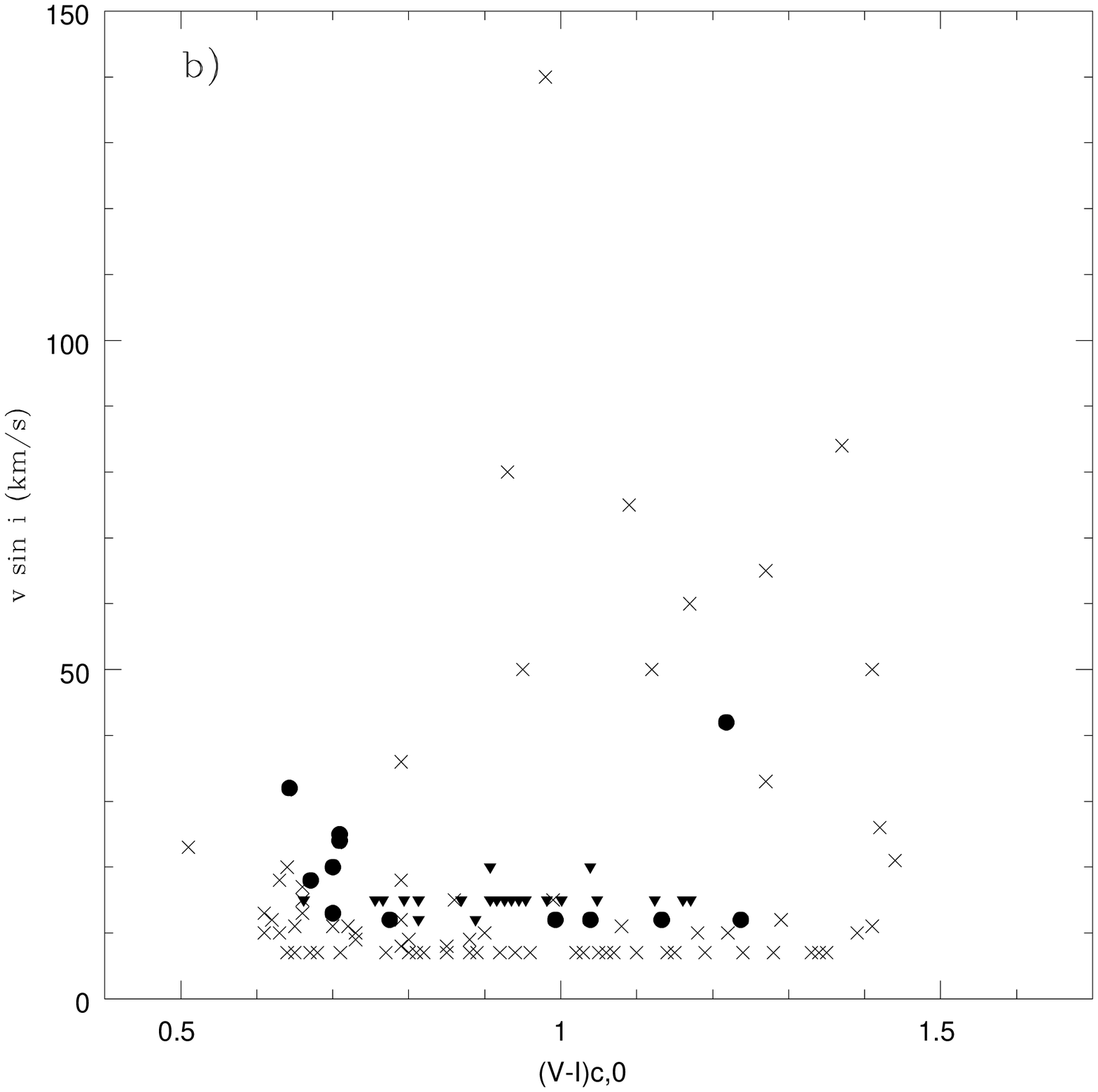}
\caption{{\bf b}}
\end{figure*}
%%%%%%%%%%%%%%%%%%%%%%%%%%%%%%%%

%%%%%%%%%%%%%%%%%%%%%%%%%%%%%%%%
\setcounter{figure}{2}
\begin{figure*}
\vspace{18cm}
\includegraphics{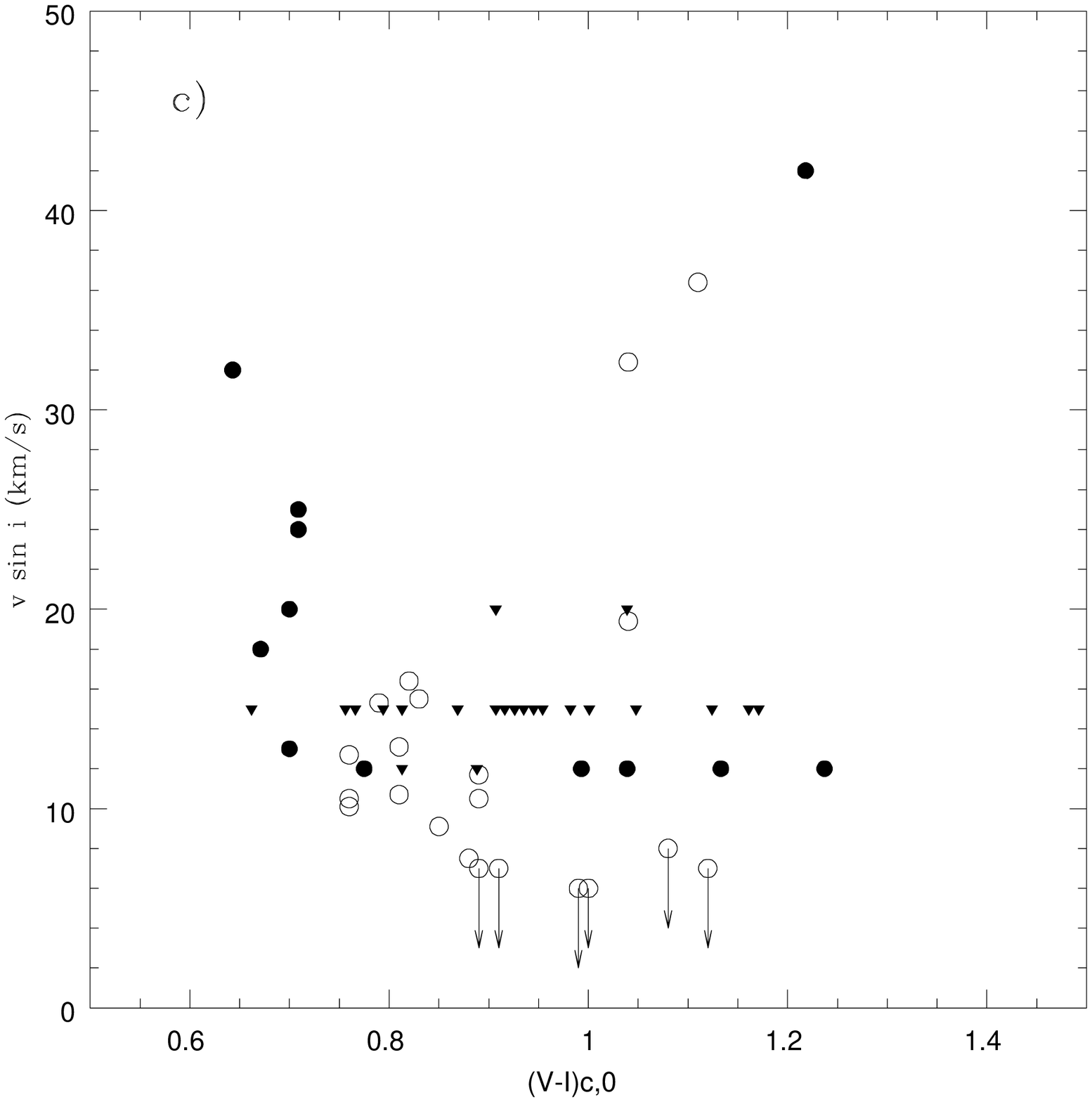}
\caption{{\bf c} }
\end{figure*}
%%%%%%%%%%%%%%%%%%%%%%%%%%%%%%%%

%%%%%%%%%%%%%%%%%%%%%%%%%%%%%%%%
\begin{figure*}
\vspace{18cm}
\includegraphics{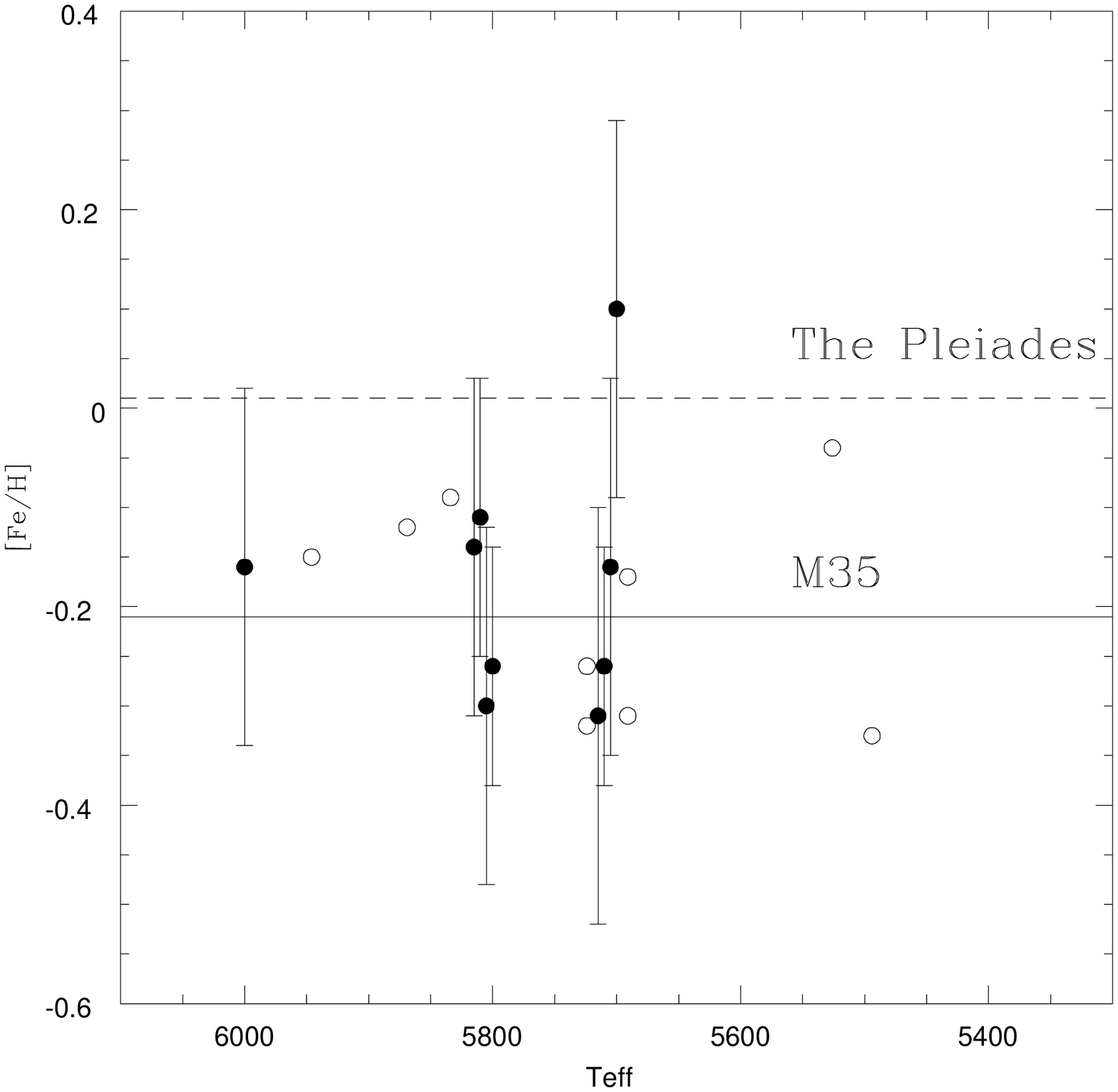}
\caption{}
\end{figure*}
%%%%%%%%%%%%%%%%%%%%%%%%%%%%%%%%

%%%%%%%%%%%%%%%%%%%%%%%%%%%%%%%%
\begin{figure*}
\vspace{18cm}
\includegraphics{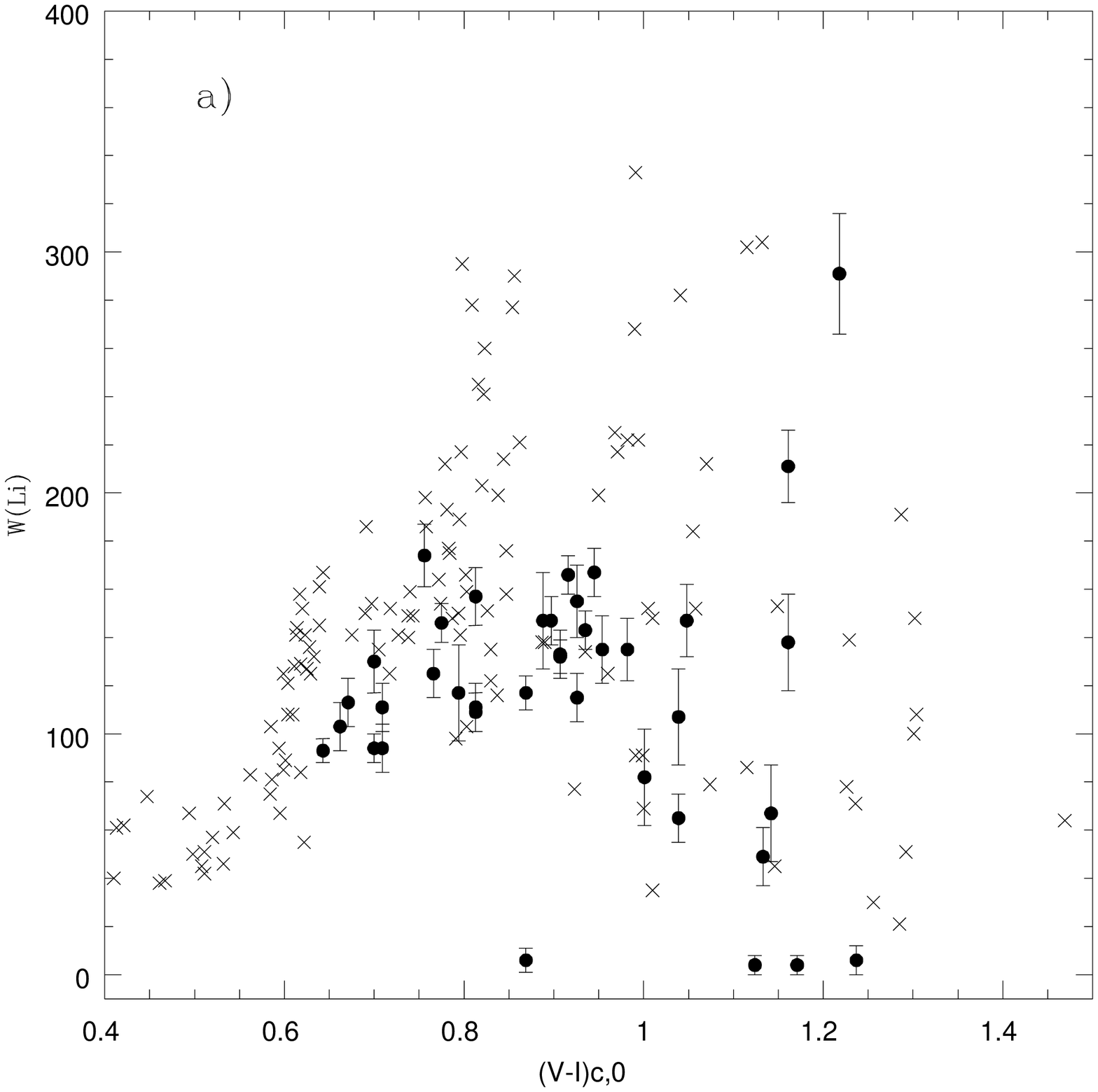}
\caption{{\bf a} }
\end{figure*}
%%%%%%%%%%%%%%%%%%%%%%%%%%%%%%%%

%%%%%%%%%%%%%%%%%%%%%%%%%%%%%%%%
\setcounter{figure}{4}
\begin{figure*}
\vspace{18cm}
\includegraphics{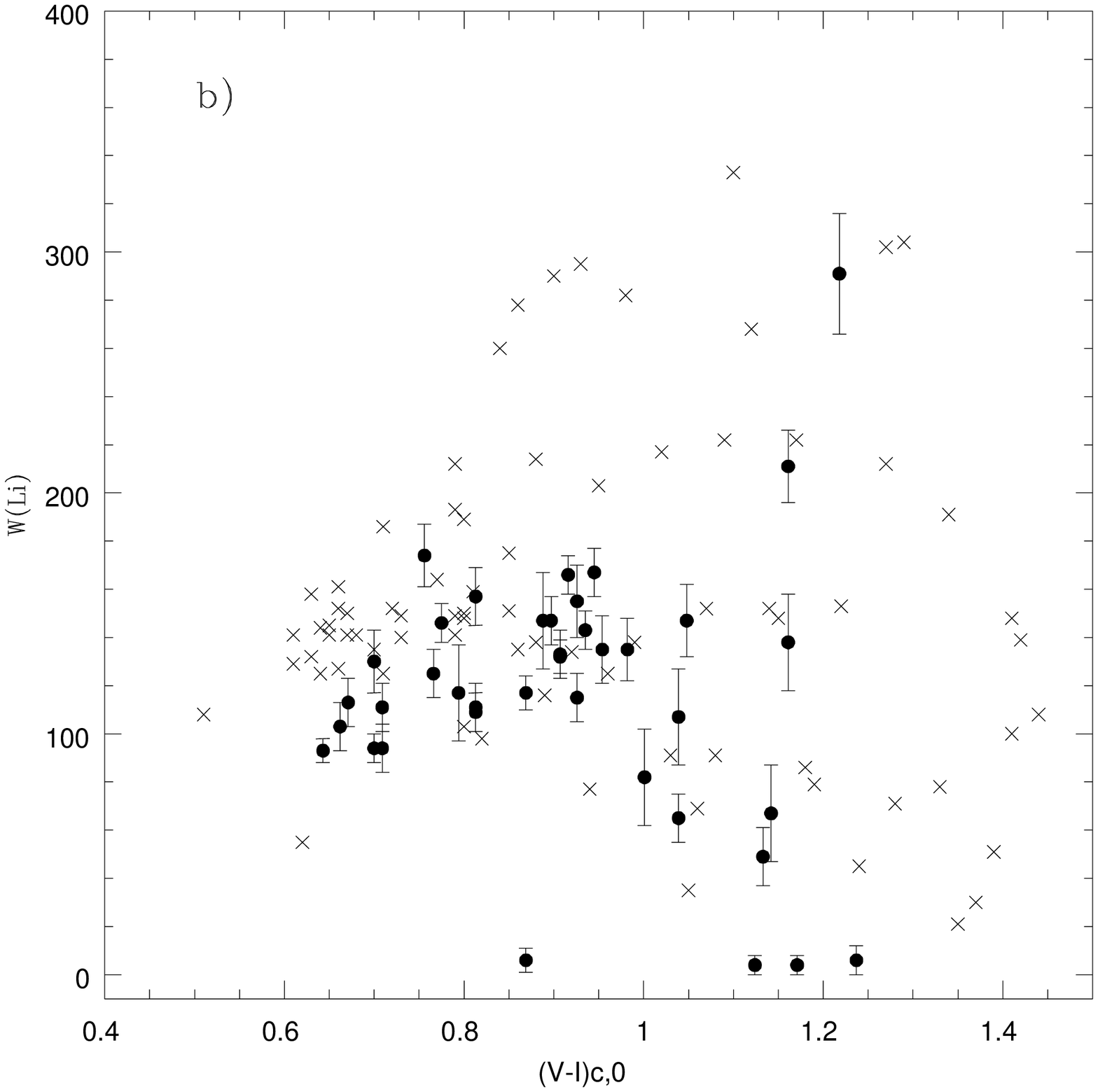}
\caption{{\bf b}}
\end{figure*}
%%%%%%%%%%%%%%%%%%%%%%%%%%%%%%%%

%%%%%%%%%%%%%%%%%%%%%%%%%%%%%%%%
\setcounter{figure}{5}
\begin{figure*}
\vspace{18cm}
\includegraphics{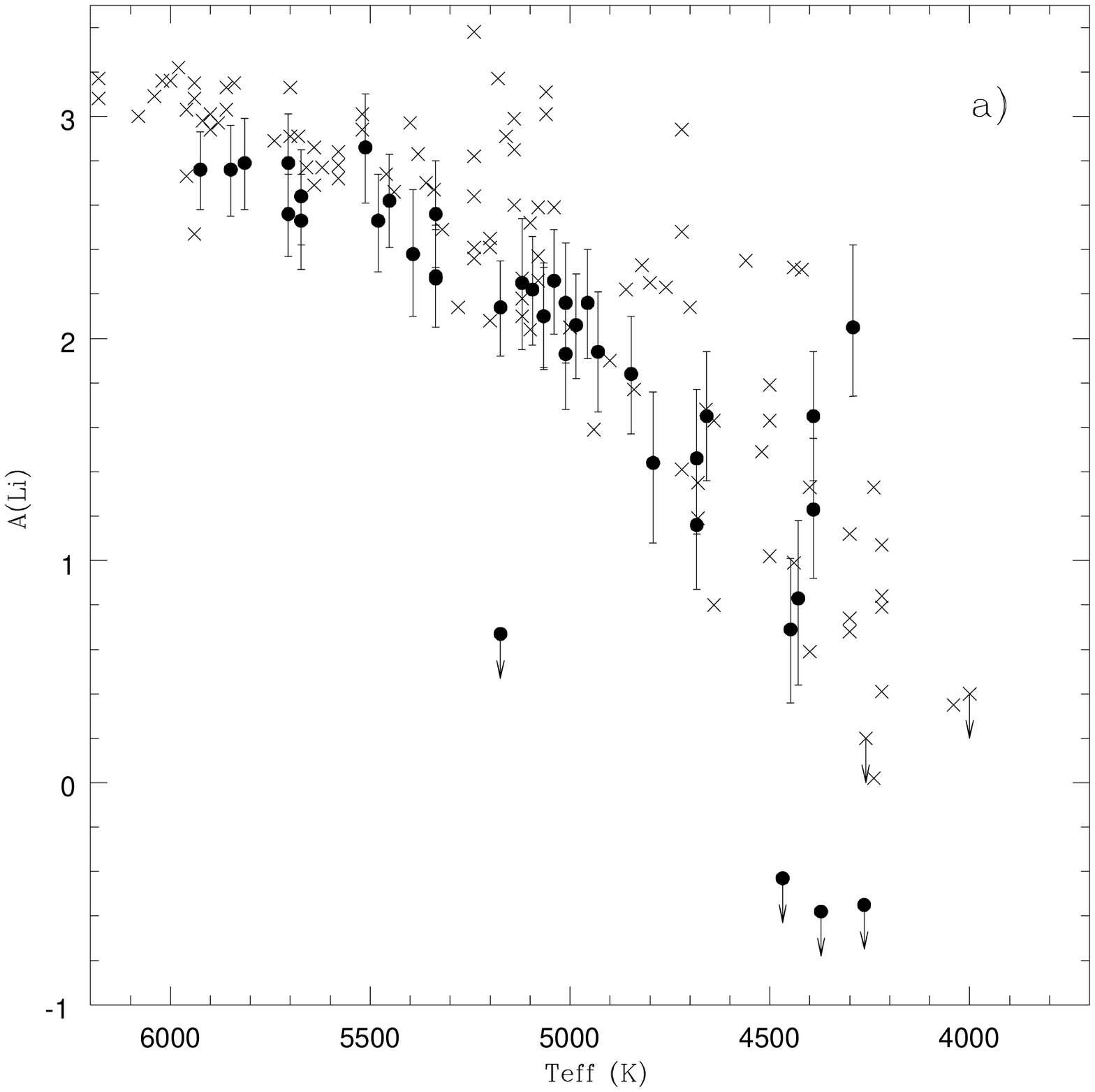}
\caption{{\bf a} }
\end{figure*}
%%%%%%%%%%%%%%%%%%%%%%%%%%%%%%%%

%%%%%%%%%%%%%%%%%%%%%%%%%%%%%%%%
\setcounter{figure}{5}
\begin{figure*}
\vspace{18cm}
\includegraphics{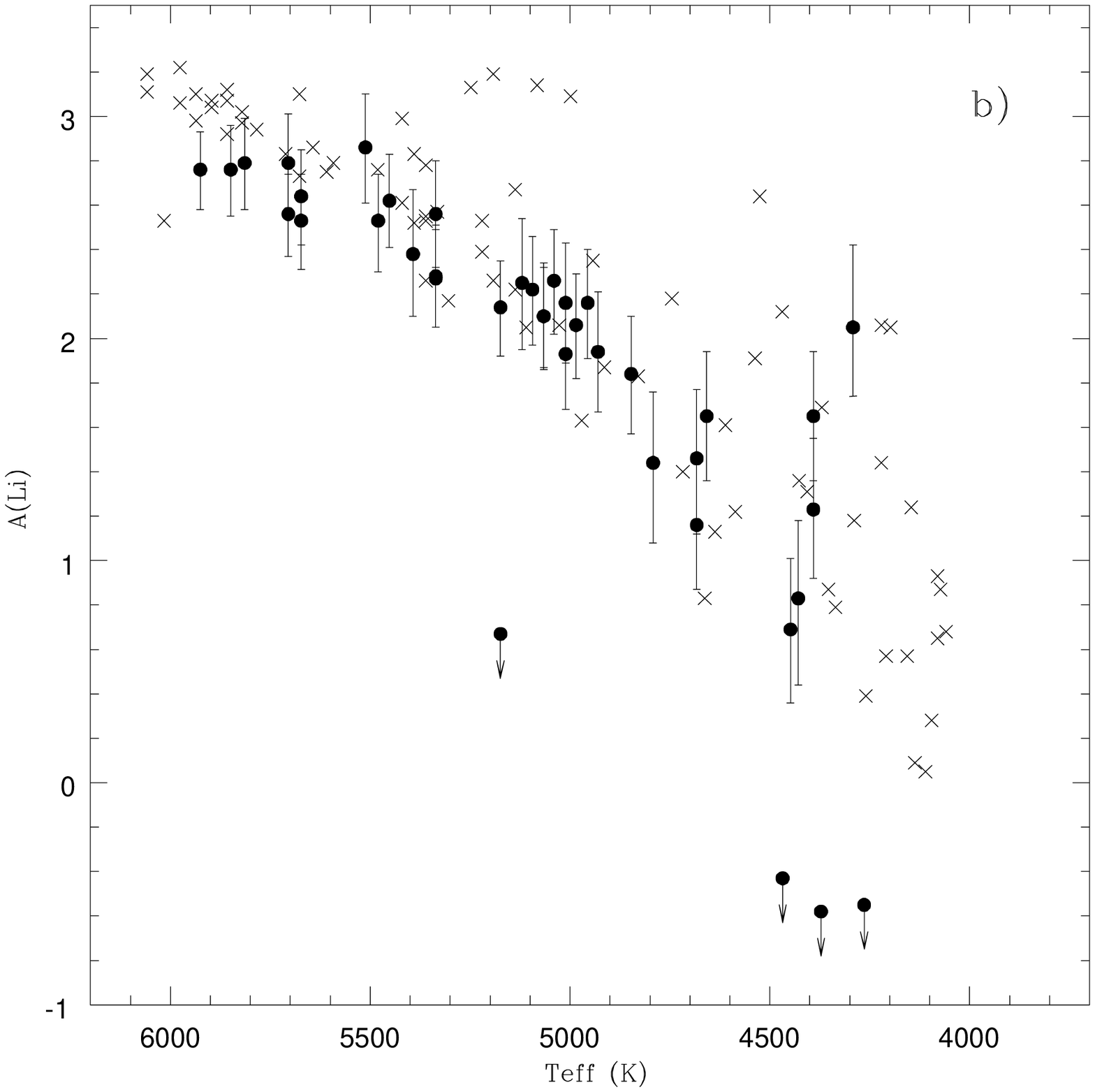}
\caption{{\bf b} }
\end{figure*}
%%%%%%%%%%%%%%%%%%%%%%%%%%%%%%%%

%%%%%%%%%%%%%%%%%%%%%%%%%%%%%%%%
\setcounter{figure}{6}
\begin{figure*}
\vspace{18cm}
\includegraphics{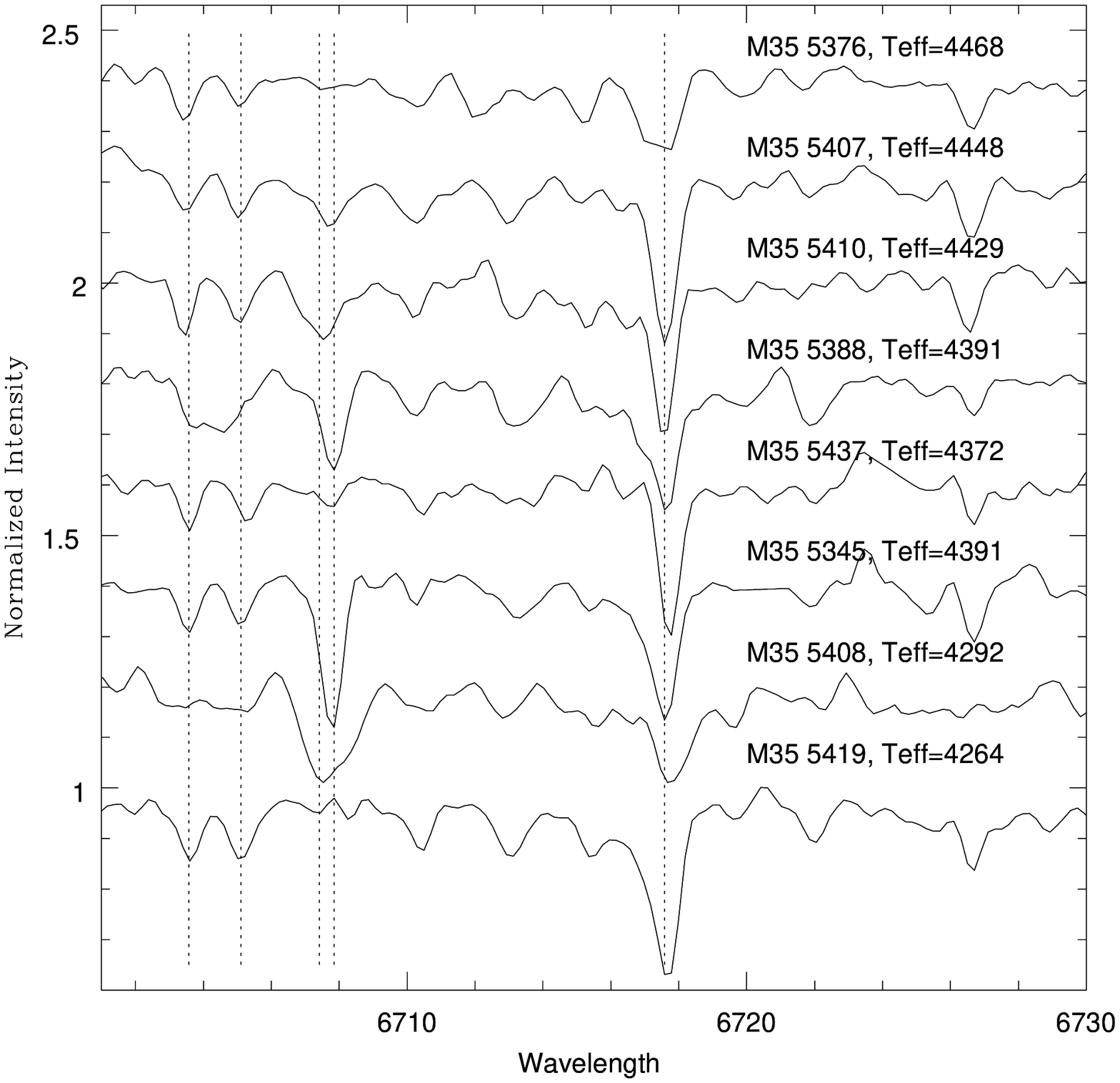}
\caption{ }
\end{figure*}
%%%%%%%%%%%%%%%%%%%%%%%%%%%%%%%%

%%%%%%%%%%%%%%%%%%%%%%%%%%%%%%%%
\setcounter{figure}{7}
\begin{figure*}
\vspace{18cm}
\includegraphics{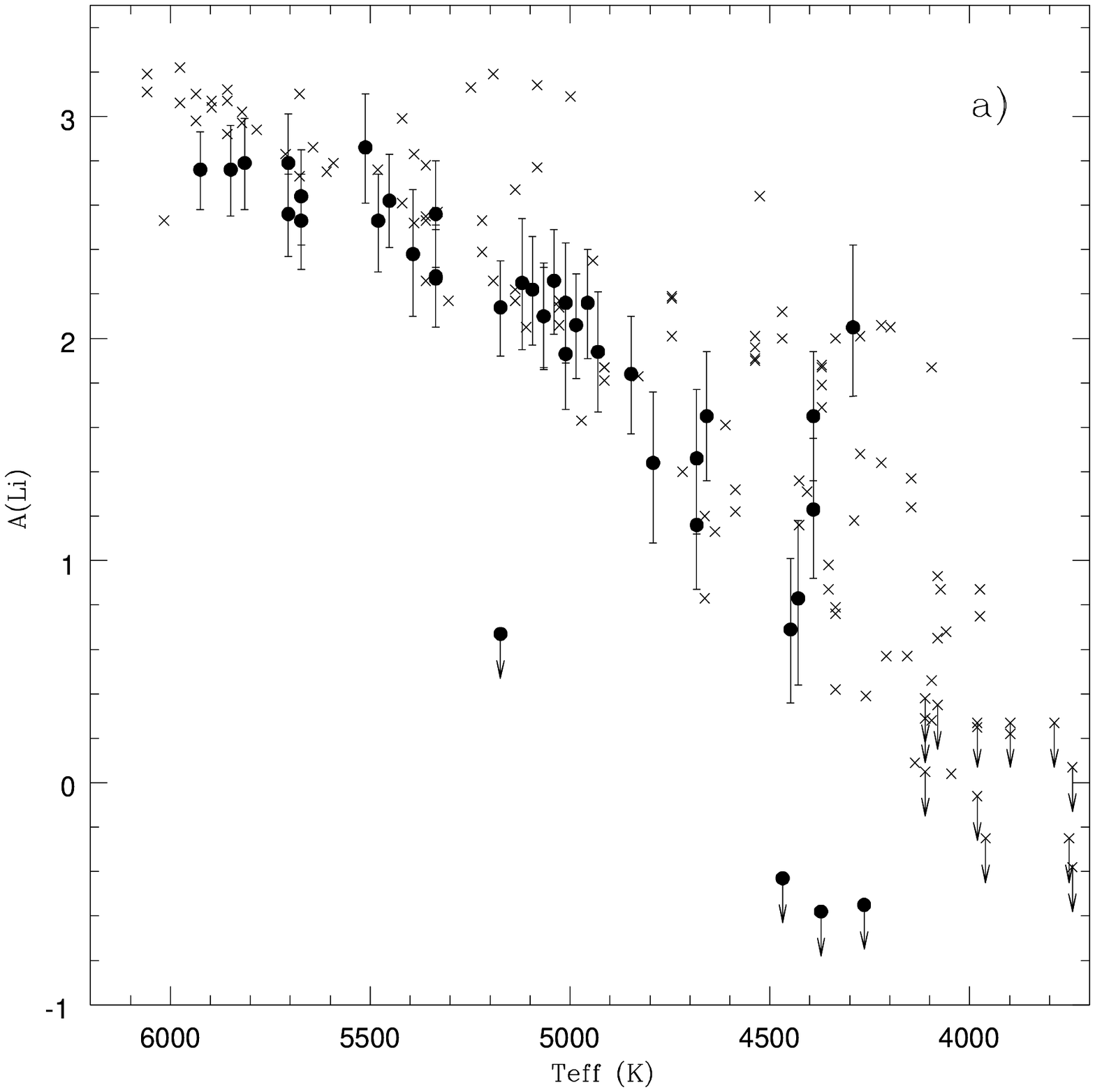}
\caption{{\bf a}  }
\end{figure*}
%%%%%%%%%%%%%%%%%%%%%%%%%%%%%%%%

%%%%%%%%%%%%%%%%%%%%%%%%%%%%%%%%
\setcounter{figure}{7}
\begin{figure*}
\vspace{18cm}
\includegraphics{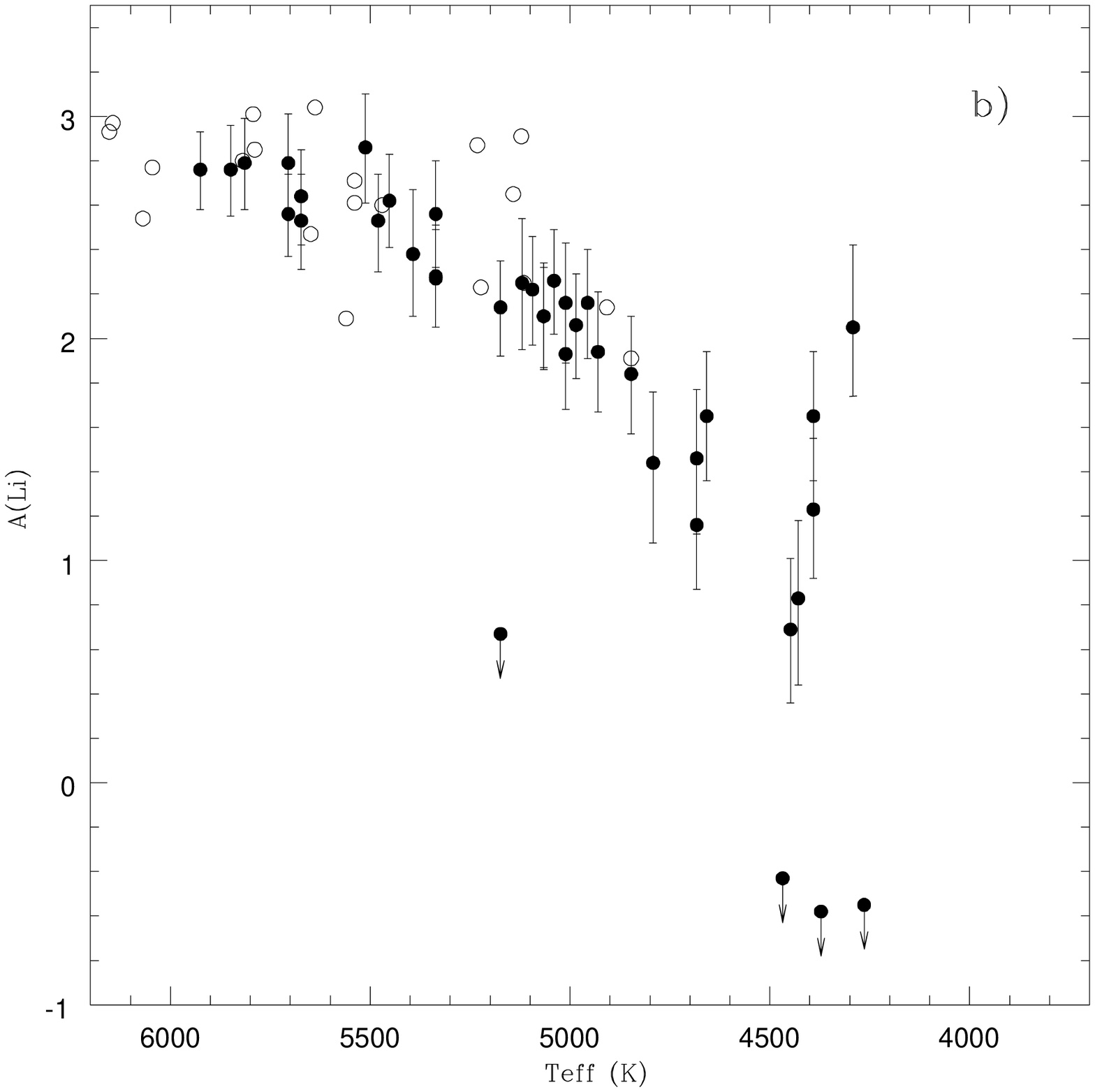}
\caption{{\bf b} }
\end{figure*}
%%%%%%%%%%%%%%%%%%%%%%%%%%%%%%%%

%%%%%%%%%%%%%%%%%%%%%%%%%%%%%%%%
\setcounter{figure}{7}
\begin{figure*}
\vspace{18cm}
\includegraphics{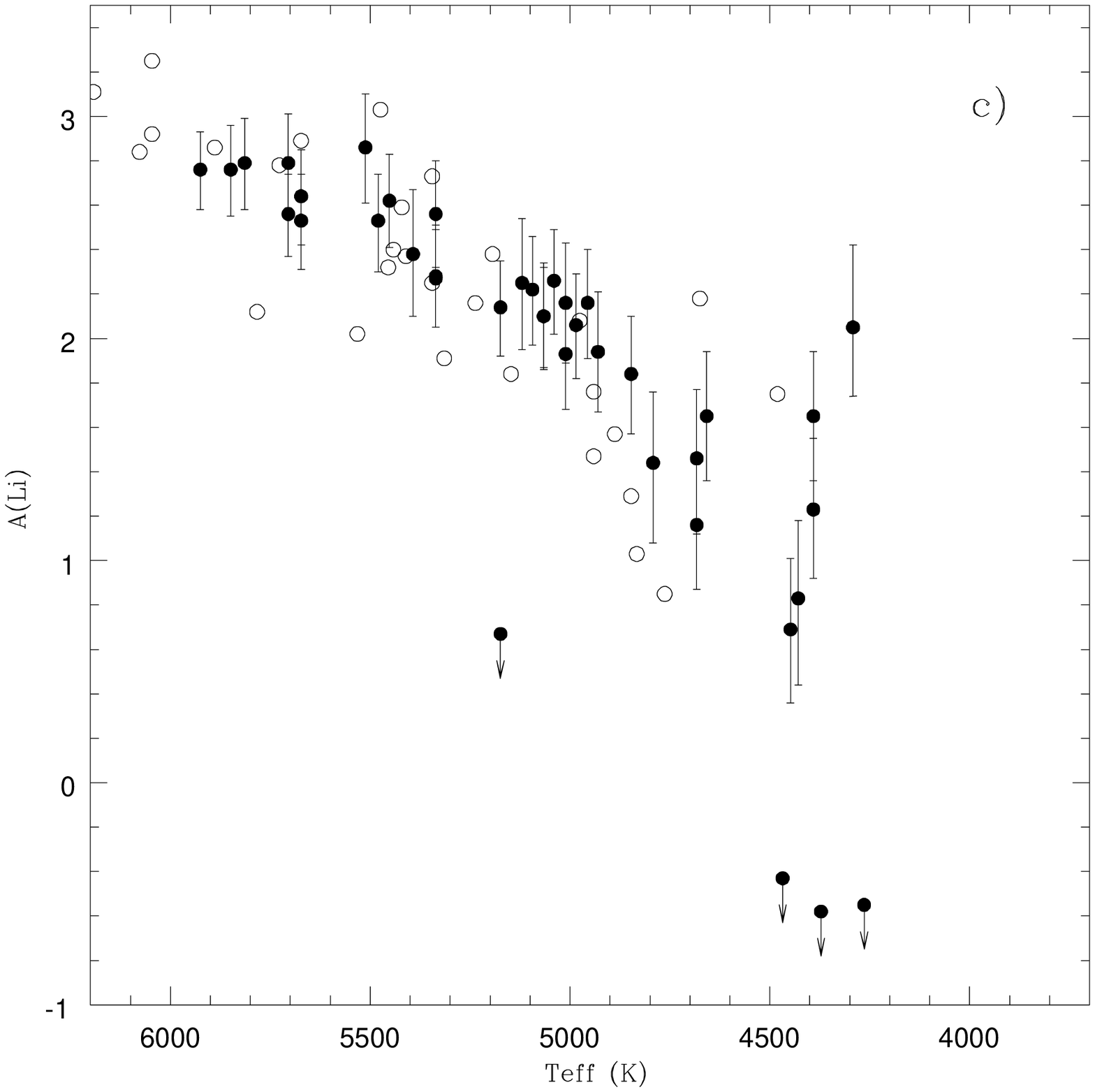}
\caption{{\bf c} }
\end{figure*}
%%%%%%%%%%%%%%%%%%%%%%%%%%%%%%%%

%%%%%%%%%%%%%%%%%%%%%%%%%%%%%%%%
\setcounter{figure}{7}
\begin{figure*}
\vspace{18cm}
\includegraphics{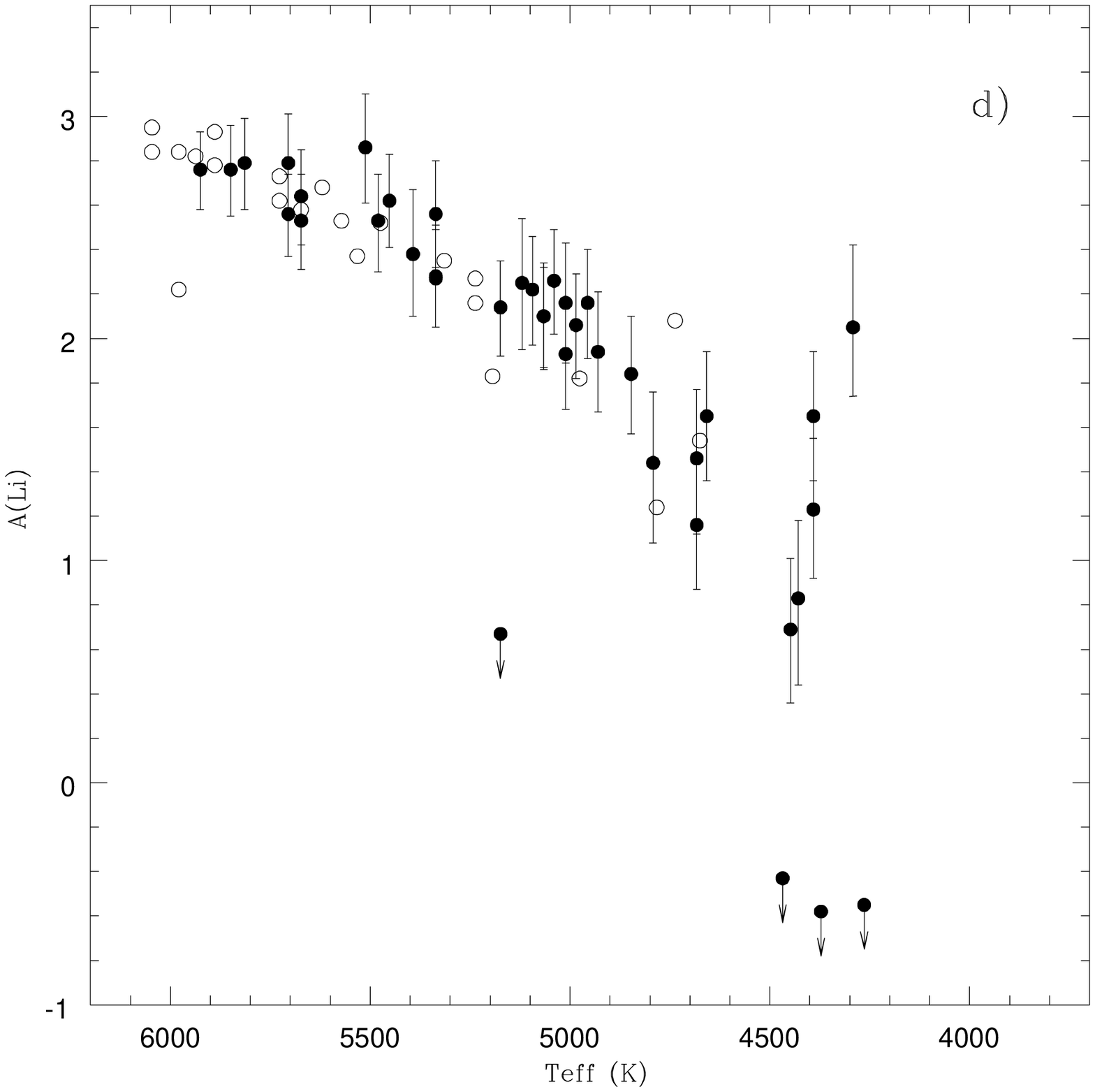}
\caption{{\bf d}}
\end{figure*}
%%%%%%%%%%%%%%%%%%%%%%%%%%%%%%%%

%%%%%%%%%%%%%%%%%%%%%%%%%%%%%%%%
\setcounter{figure}{7}
\begin{figure*}
\vspace{18cm}
\includegraphics{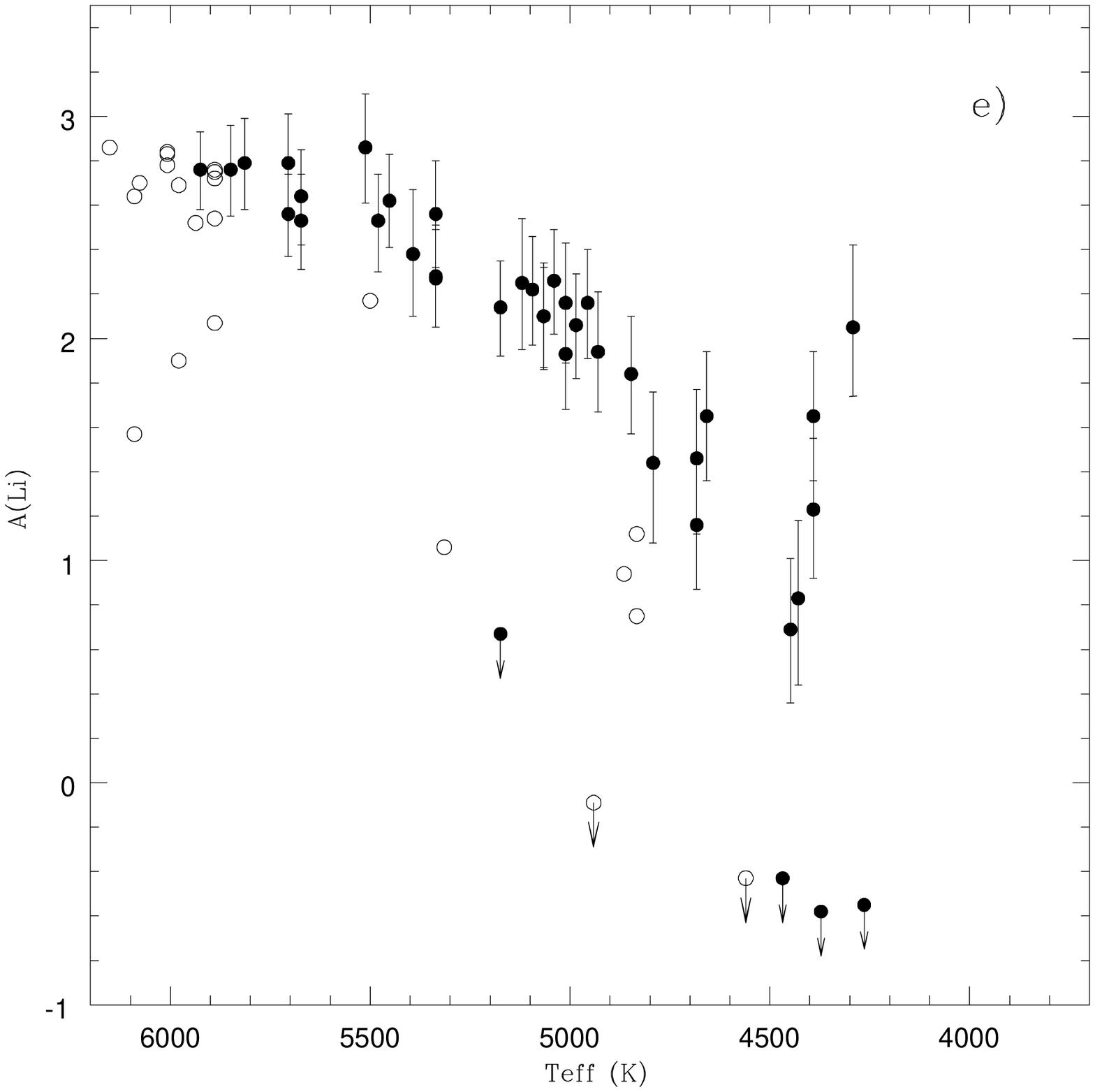}
\caption{{\bf e} }
\end{figure*}
%%%%%%%%%%%%%%%%%%%%%%%%%%%%%%%%

%%%%%%%%%%%%%%%%%%%%%%%%%%%%%%%%
\setcounter{figure}{7}
\begin{figure*}
\vspace{18cm}
\includegraphics{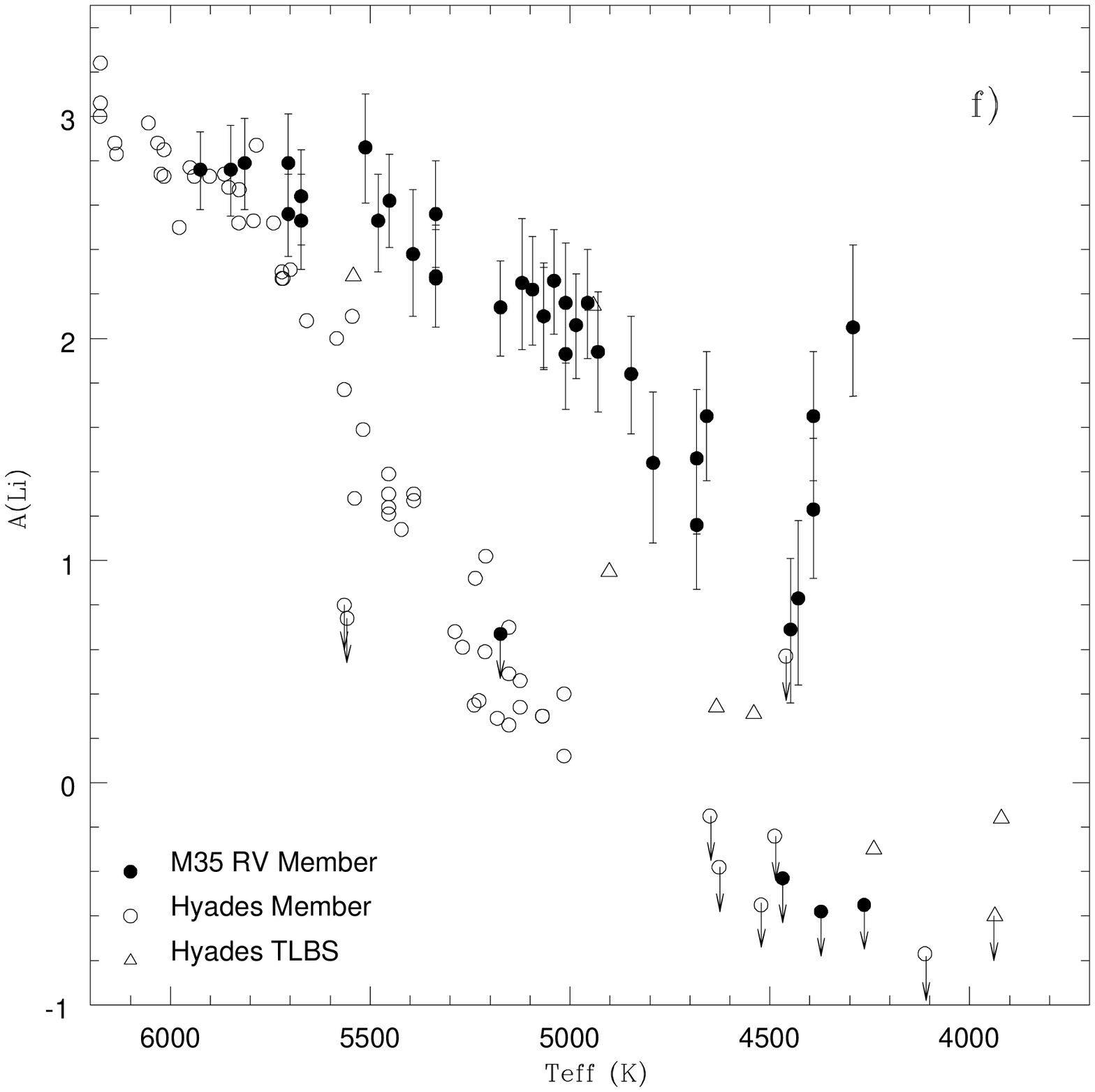}
\caption{{\bf f} }
\end{figure*}
%%%%%%%%%%%%%%%%%%%%%%%%%%%%%%%%

\end{document}

\baselineskip=16pt
\pagestyle{empty}

\normalsize

\footnotesize

\setcounter{table}{0}
 
\begin{table}
\caption[ ]{Positions, photometric data and membership status  for our M35 candidate members.}
\begin{tabular}{lcccccccccccl}
\hline
ID & Cuffey/&   alpha      &    delta   &&V&(V-I)$_C$&&  V    &(V-I)$_C$& (B-V)  & (U-B) & Member?    \\  %     KPNO_old
\cline{3-4}\cline{6-7}\cline{9-12}  
  \#  &  WEBDA   &\multicolumn{2}{c}{(2000.0)}&&\multicolumn{2}{c}{ByN2000}&& \multicolumn{4}{c}{Sung \& Bessel (1999)} &   \\  % 
(1) &  (2)&  (3) &     (4)    &  &   (5)  &  (6)  &  &  (7)  &    (8)  &  (9)   &  (10)    &  (11)   \\  %  		      
\hline									  
5081 & 0691 &  6:09:26.200 & 24:29:03.0 &\,& 14.628 & 0.951 &\,&  14.879 &  1.013 &   0.768 &  0.313  &  SB1     \\  %   1079  
5089 &  --  &  6:09:10.230 & 24:31:44.8 &\,& 14.622 & 0.857 &\,&   --    &  --    &   --    &  --     &  MEM     \\  %   1169  
5096 & 1030 &  6:08:19.030 & 24:28:37.1 &\,& 14.672 & 0.876 &\,&  14.657 &  0.872 &   0.737 &  0.173  &  MEM     \\  %   1393  
5098 & 0178 &  6:09:26.710 & 24:18:07.2 &\,& 14.720 & 0.914 &\,&  14.720 &  0.913 &   0.751 &  0.170  &  MEM     \\  %   1082  
5103 &  --  &  6:09:19.600 & 24:31:57.6 &\,& 14.741 & 0.885 &\,&   --    &  --    &   --    &  --     &  MEM     \\  %   1119  
5107 & 0032 &  6:08:58.860 & 24:20:17.2 &\,& 14.809 & 0.923 &\,&  14.859 &  0.936 &   0.786 &  0.169  &  MEM     \\  %   1225  
5113 & 0405 &  6:08:47.580 & 24:12:04.2 &\,& 14.868 & 0.951 &\,&  14.863 &  0.957 &   0.787 &  0.225  &  SB1     \\  %   1277  
5116 & 0251 &  6:08:40.600 & 24:15:58.4 &\,& 14.859 & 0.923 &\,&  14.877 &  0.942 &   0.771 &  0.199  &  MEM     \\  %   1309  
5125 & 3856 &  6:09:14.810 & 24:31:23.7 &\,& 14.920 & 0.914 &\,&  15.234 &  0.984 &   0.775 &  0.221  &  MEM     \\  %   1144    
5126 & 0704 &  6:09:36.270 & 24:27:08.9 &\,& 15.016 & 0.998 &\,&  15.156 &  1.050 &   0.857 &  0.324  &  NM      \\  %   1042  
5132 & 0037 &  6:08:59.070 & 24:21:00.4 &\,& 15.058 & 0.942 &\,&  15.081 &  0.985 &   0.776 &  0.352  &  SB2     \\  %   1222  
5135 & 0128 &  6:08:52.220 & 24:23:28.6 &\,& 15.126 & 0.980 &\,&  15.195 &  1.030 &   0.831 &  0.265  &  MEM     \\  %   1254  
5136 & 0004 &  6:09:09.970 & 24:18:54.5 &\,& 15.118 & 0.951 &\,&  15.141 &  0.960 &   0.748 &  0.307  &  SB2     \\  %   1174  
5138 & 3075 &  6:08:15.410 & 24:25:27.8 &\,& 15.186 & 0.989 &\,&  15.194 &  1.003 &   0.824 &  0.261  &  MEM     \\  %   1408  
5141 &  --  &  6:08:48.550 & 24:10:29.0 &\,& 15.236 & 0.998 &\,&   --    &  --    &   --    &  --     &  NM      \\  %   1268  
5142 &  --  &  6:09:43.690 & 24:21:04.5 &\,& 15.216 & 0.989 &\,&   --    &  --    &   --    &  --     &  NM      \\  %   1008  
5144 & 0182 &  6:09:28.520 & 24:15:32.1 &\,& 15.264 & 1.027 &\,&  15.327 &  1.056 &   0.866 &  0.347  &  MEM     \\  %   1074  
5151 & 0314 &  6:09:16.980 & 24:24:19.5 &\,& 15.296 & 0.980 &\,&  15.327 &  1.021 &   --    &  --     &  SB1     \\  %   1136  
5159 & 0354 &  6:09:33.840 & 24:14:59.9 &\,& 15.511 & 1.093 &\,&  15.525 &  1.115 &   0.899 &  0.309  &  NM      \\  %   1051  
5165 & 0383 &  6:09:15.120 & 24:11:04.1 &\,& 15.464 & 1.027 &\,&  15.486 &  1.052 &   0.846 &  0.345  &  MEM     \\  %   1146  
5174 & 3015 &  6:08:10.430 & 24:24:24.7 &\,& 15.467 & 0.970 &\,&  15.452 &  0.982 &   0.838 &  0.334  &  MEM     \\  %   1418  
5176 & 3597 &  6:08:56.400 & 24:27:10.9 &\,& 15.505 & 1.008 &\,&  15.562 &  1.027 &   0.903 &  0.356  &  MEM     \\  %   1232  
5178 & 0048 &  6:09:10.840 & 24:19:46.3 &\,& 15.506 & 0.989 &\,&  15.544 &  1.026 &   0.811 &  0.239  &  NM      \\  %   1171  
5184 & 0167 &  6:09:20.740 & 24:19:55.2 &\,& 15.536 & 0.998 &\,&  15.585 &  1.051 &   0.794 &  0.177  &  NM      \\  %   1116  
5190 &  --  &  6:09:16.060 & 24:25:31.9 &\,& 15.671 & 1.083 &\,&   --    &  --    &   --    &  --     &  MEM     \\  %   1443  
5192 & 0606 &  6:08:24.170 & 24:12:23.2 &\,& 15.605 & 1.008 &\,&  15.619 &  1.006 &   0.792 &  0.181  &  NM      \\  %   1383  
5194 & 0017 &  6:09:03.040 & 24:19:36.8 &\,& 15.758 & 1.149 &\,&  15.768 &  1.169 &   0.949 &  0.422  &  SB1     \\  %   1212  
5195 & 3456 &  6:08:45.830 & 24:18:23.0 &\,& 15.740 & 1.102 &\,&  15.776 &  1.141 &   0.913 &  0.357  &  NM      \\  %   1284  
5205 & 0126 &  6:08:54.800 & 24:22:51.1 &\,& 15.780 & 1.102 &\,&  15.811 &  1.142 &   1.009 &  0.548  &  NM      \\  %   1242  
5209 & 3088 &  6:08:17.040 & 24:27:02.6 &\,& 15.734 & 1.027 &\,&  15.742 &  1.041 &   0.919 &  0.436  &  MEM     \\  %   1401  
5216 & 3279 &  6:08:32.690 & 24:27:18.9 &\,& 15.754 & 1.027 &\,&  15.756 &  1.058 &   0.910 &  0.385  &  NM      \\  %   1343  
5219 & 0289 &  6:08:47.330 & 24:24:56.1 &\,& 15.803 & 1.046 &\,&  15.826 &  1.090 &   0.999 &  0.523  &  NM      \\  %   1274  
5231 & 3538 &  6:08:52.710 & 24:24:44.9 &\,& 15.920 & 1.121 &\,&  15.994 &  1.182 &   0.989 &  0.458  &  NM      \\  %   1251  
5233 & 0387 &  6:09:11.690 & 24:11:24.5 &\,& 15.930 & 1.121 &\,&  15.942 &  1.108 &   0.939 &  0.542  &  MEM     \\  %   1164  
5235 & 0026 &  6:08:59.030 & 24:18:25.6 &\,& 15.968 & 1.149 &\,&  16.016 &  1.172 &   1.014 &  0.538  &  MEM     \\  %   1224  
5236 &  --  &  6:09:42.910 & 24:31:45.0 &\,& 15.901 & 1.083 &\,&   --    &  --    &   --    &  --     &  NM      \\  %   1009  
5241 & 0055 &  6:09:21.840 & 24:18:22.0 &\,& 15.930 & 1.102 &\,&  15.980 &  1.128 &   0.977 &  0.503  &  MEM     \\  %   1111  
5243 &  --  &  6:09:09.410 & 24:10:45.7 &\,& 15.902 & 1.064 &\,&   --    &  --    &   --    &  --     &  SB1     \\  %   1180   
5257 & 0011 &  6:09:06.460 & 24:18:00.3 &\,& 16.001 & 1.083 &\,&  16.048 &  1.150 &   1.000 &  0.495  &  MEM     \\  %   1194  
5263 & 3843 &  6:09:14.070 & 24:13:37.5 &\,& 16.070 & 1.111 &\,&  16.104 &  1.137 &   0.999 &  0.559  &  MEM     \\  %   1149  
5264 & 3751 &  6:09:07.260 & 24:23:45.0 &\,& 16.080 & 1.121 &\,&  16.139 &  1.191 &   0.985 &  0.431  &  NM      \\  %   1188  
5267 &  --  &  6:08:57.650 & 24:10:01.3 &\,& 16.109 & 1.140 &\,&   --    &  --    &   --    &  --     &  MEM     \\  %   1231  
5268 & 0333 &  6:09:26.480 & 24:20:47.2 &\,& 16.061 & 1.083 &\,&  16.099 &  1.134 &   0.849 &  0.220  &  SB1     \\  %   1083  
5274 & 4093 &  6:09:32.970 & 24:25:18.5 &\,& 16.139 & 1.130 &\,&  16.215 &  1.179 &   1.010 &  0.678  &  MEM     \\  %   1052  
5288 & 4040 &  6:09:27.930 & 24:22:27.4 &\,& 16.259 & 1.140 &\,&  16.293 &  1.207 &   1.066 &  0.602  &  MEM     \\  %   1076  
5289 & 4050 &  6:09:29.170 & 24:21:52.1 &\,& 16.240 & 1.121 &\,&  16.313 &  1.222 &   1.040 &  0.674  &  MEM     \\  %   1072  
5292 & 3835 &  6:09:13.470 & 24:21:46.8 &\,& 16.374 & 1.243 &\,&  16.444 &  1.325 &   1.108 &  0.622  &  NM      \\  %   1151  
5293 & 3743 &  6:09:06.460 & 24:21:30.8 &\,& 16.336 & 1.196 &\,&  16.374 &  1.223 &   1.051 &  0.590  &  MEM     \\  %   1193  
5294 & 3493 &  6:08:48.750 & 24:16:26.6 &\,& 16.308 & 1.159 &\,&  16.343 &  1.195 &   0.919 &  0.216  &  NM      \\  %   1266  
5297 &  --  &  6:08:33.880 & 24:09:48.4 &\,& 16.338 & 1.159 &\,&   --    &  --    &   --    &  --     &  MEM     \\  %   1341  
5299 & 3505 &  6:08:49.620 & 24:14:40.1 &\,& 16.424 & 1.243 &\,&  16.480 &  1.295 &   1.116 &  0.944  &  NM      \\  %   1263  
\hline
\end{tabular}
$\,$\\  % 
ID numbers (column 1) and photometry (columns 5 \& 6) from  from Barrado y Navascu\'es et al. (2000b)\\
WEBDA numbers (Open Cluster database,  Mermilliod  1996) 
coincide with the identification numbers\\
 by Cuffey (1938) for the first 778 stars in the WEBDA database.\\    %
MEM = Probable Member; NM = Probable non-member;  SB = Spectroscopic binary, possible member.\\  %
\end{table}

\setcounter{table}{0}

\begin{table}
\caption[ ]{Positions, photometric data and membership status  for our M35 candidate members (cont.).}
\begin{tabular}{lcccccccccccl}
\hline
ID & Cuffey/&   alpha      &    delta   &&V&(V-I)$_C$&&  V    &(V-I)$_C$& (B-V)  & (U-B) & Member?    \\  %     KPNO_old
\cline{3-4}\cline{6-7}\cline{9-12}  
  \#  &  WEBDA   &\multicolumn{2}{c}{(2000.0)}&&\multicolumn{2}{c}{ByN2000}&& \multicolumn{4}{c}{Sung \& Bessel (1999)} &   \\  % 
(1) &  (2)&  (3) &     (4)    &  &   (5)  &  (6)  &  &  (7)  &    (8)  &  (9)   &  (10)    &  (11)   \\  %  		      
\hline									  
5305 & 3365 &  6:08:39.120 & 24:24:19.7 &\,& 16.340 & 1.121 &\,&  16.348 &  1.158 &   0.956 &  0.387  &  NM      \\  %   1316  
5309 &  --  &  6:08:06.140 & 24:27:19.8 &\,& 16.493 & 1.253 &\,&   --    &  --    &   --    &  --     &  NM      \\  %   1431  
5314 &  --  &  6:08:39.200 & 24:10:45.8 &\,& 16.476 & 1.196 &\,&   --    &  --    &   --    &  --     &  NM      \\  %   1318  
5316 & 4024 &  6:09:26.760 & 24:27:55.5 &\,& 16.447 & 1.168 &\,&  16.588 &  1.209 &   1.064 &  0.884  &  MEM     \\  %   1078  
5317 & 3145 &  6:08:22.260 & 24:15:03.3 &\,& 16.505 & 1.215 &\,&  16.521 &  1.185 &   1.083 &  0.772  &  MEM     \\  %   1388  
5332 & 3591 &  6:08:55.970 & 24:25:11.5 &\,& 16.693 & 1.262 &\,&  16.789 &  1.348 &   1.151 &  0.708  &  MEM     \\  %   1237  
5339 & 3880 &  6:09:16.290 & 24:18:37.3 &\,& 16.723 & 1.253 &\,&  16.751 &  1.266 &   1.138 &  1.041  &  MEM     \\  %   1141  
5345 & 3623 &  6:08:58.330 & 24:14:22.2 &\,& 16.897 & 1.375 &\,&  16.851 &  1.398 &   1.162 &  0.782  &  MEM     \\  %   1229  
5347 & 3936 &  6:09:20.340 & 24:23:05.2 &\,& 16.716 & 1.196 &\,&  16.766 &  1.241 &   1.104 &  0.721  &  NM      \\  %   1117  
5350 &  --  &  6:09:41.770 & 24:14:59.9 &\,& 16.726 & 1.196 &\,&   --    &  --    &   --    &  --     &  SB2     \\  %   1017  
5357 & 3384 &  6:08:40.610 & 24:25:40.4 &\,& 16.803 & 1.253 &\,&  16.802 &  1.294 &   1.164 &  0.894  &  MEM     \\  %   1303  
5376 &  --  &  6:09:34.510 & 24:23:53.9 &\,& 16.969 & 1.338 &\,&   --    &  --    &   --    &  --     &  MEM     \\  %   1047  
5384 &  --  &  6:09:06.130 & 24:27:20.4 &\,& 17.095 & 1.422 &\,&   --    &  --    &   --    &  --     &  NM      \\  %   1192  
5388 &  --  &  6:08:48.250 & 24:17:59.8 &\,& 17.077 & 1.375 &\,&   --    &  --    &   --    &  --     &  MEM     \\  %   1267  
5398 &  --  &  6:08:52.550 & 24:22:00.6 &\,& 17.166 & 1.394 &\,&   --    &  --    &   --    &  --     &  SB1     \\  %   1253  
5399 &  --  &  6:09:12.600 & 24:22:24.6 &\,& 17.224 & 1.441 &\,&   --    &  --    &   --    &  --     &  SB1     \\  %   1156  
5407 &  --  &  6:09:17.820 & 24:21:57.1 &\,& 17.169 & 1.347 &\,&   --    &  --    &   --    &  --     &  MEM     \\  %   1133  
5408 &  --  &  6:08:15.230 & 24:28:06.1 &\,& 17.254 & 1.432 &\,&   --    &  --    &   --    &  --     &  MEM     \\  %   1407  
5410 &  --  &  6:08:23.370 & 24:23:55.9 &\,& 17.188 & 1.356 &\,&   --    &  --    &   --    &  --     &  MEM     \\  %   1385  
5411 &  --  &  6:08:37.390 & 24:20:44.0 &\,& 17.331 & 1.498 &\,&   --    &  --    &   --    &  --     &  NM      \\  %   1322  
5419 &  --  &  6:08:43.410 & 24:20:28.5 &\,& 17.333 & 1.451 &\,&   --    &  --    &   --    &  --     &  MEM     \\  %   1291  
5432 &  --  &  6:09:23.670 & 24:31:56.3 &\,& 17.538 & 1.564 &\,&   --    &  --    &   --    &  --     &  NM      \\  %   1096  
5437 &  --  &  6:09:23.580 & 24:26:19.2 &\,& 17.377 & 1.385 &\,&   --    &  --    &   --    &  --     &  MEM     \\  %   1099  
5449 &  --  &  6:09:24.830 & 24:14:26.5 &\,& 17.532 & 1.488 &\,&   --    &  --    &   --    &  --     &  SB1     \\  %   1095  
5454 &  --  &  6:08:27.400 & 24:12:08.6 &\,& 17.649 & 1.535 &\,&   --    &  --    &   --    &  --     &  SB1     \\  %   1369  
\hline
\end{tabular}$\,$\\  % 
ID numbers (column 1) and photometry (columns 5 \& 6) from  from Barrado y Navascu\'es et al. (2000b)\\
WEBDA numbers (Mermilliod database) coincide with the identification numbers\\
 by Cuffey (1938) for the first 778 stars in the WEBDA database.\\    %
MEM = Probable Member; NM = Probable non-member;  SB = Spectroscopic binary, possible member.\\  %
\end{table}

\footnotesize

\setcounter{table}{1}

\begin{table}
\caption[]{Lithium abundances, radial and rotational velocities  for our probable members of M35.}
\begin{tabular}{cccrrrrcrccccr}
\hline
 ID  &Vhelio&W(Li+Fe)&W(Li)&$\Delta$W(Li)&Teff  &  A(Li)&\,& Teff  & A(Li)&$\Delta$A(Li)&vsini\\%Ap KPNO Cuffey/WEBDA V-I V-Isb 
     &  (km/s)&(m\AA)  &(m\AA)&   (m\AA)   & (K)  &       &&  (K)  &        &         &(km/s)\\
     &        &        &      &            &\multicolumn{2}{c}{ByN2000}&&\multicolumn{2}{c}{SB1999}&         &      \\
\cline{6-7}\cline{9-10}
 (1) &    (2) &   (3)  &  (4) &     (5)    & (6)  &  (7)  &&  (8)  &   (9)  & (10)    &  (11)\\
\hline
%\multicolumn{12}{l}{Probable members}\\
\hline
5089 &    -8.2$\pm$2.6 &  102 & 93 & 5 &  5925 &  2.76 &\,&  --  &   --   & 0.18   &     32 \\    %   MEM  44 1169 --   0.857 & -9.99 &
5096 &    -6.8$\pm$0.2 &  113 &103 &10 &  5849 &  2.76 &\,& 5865 &  2.77  & 0.21   &$\le$15 \\    %   MEM  83 1393 1030 0.876 & 0.872 &
5098 &    -5.0$\pm$1.1 &  104 & 94 & 6 &  5705 &  2.56 &\,& 5709 &  2.56  & 0.19   &     13 \\    %   MEM  55 1082  178 0.914 & 0.913 &
5103 &    -9.8$\pm$1.6 &  123 &113 &10 &  5814 &  2.79 &\,&  --  &   --   & 0.21   &     18 \\    %   MEM  34 1119 --   0.885 & -9.99 &
5107 &   -10.5$\pm$1.2 &  104 & 94 &10 &  5673 &  2.53 &\,& 5627 &  2.48  & 0.22   &     25 \\    %   MEM  50 1225   32 0.923 & 0.936 &
5116 &    -9.3$\pm$1.8 &  121 &111 &10 &  5673 &  2.64 &\,& 5606 &  2.57  & 0.21   &     24 \\    %   MEM   4 1309  251 0.923 & 0.942 & 
5125 &    -7.0$\pm$0.8 &  140 &130 &13 &  5705 &  2.79 &\,& 5467 &  2.55  & 0.23   &     20 \\    %   MEM   6 1144 3856 0.914 & 0.984 &
5135 &    -9.0$\pm$0.4 &  136 &125 &10 &  5480 &  2.53 &\,& 5327 &  2.36  & 0.22   &$\le$15 \\    %   MEM  64 1254  128 0.980 & 1.030 &
5138 &    -9.3$\pm$0.8 &  157 &146 & 8 &  5452 &  2.62 &\,& 5408 &  2.57  & 0.21   &     12 \\    %   MEM  71 1408 3075 0.989 & 1.003 &
5144 &    -7.5$\pm$1.5 &  170 &157 &12 &  5336 &  2.56 &\,& 5251 &  2.46  & 0.24   &     12 \\    %   MEM  54 1074  182 1.027 & 1.056 &
5165 &    -5.2$\pm$2.6 &  124 &111 &10 &  5336 &  2.28 &\,& 5262 &  2.20  & 0.23   &$\le$15 \\    %   MEM  60 1146  383 1.027 & 1.052 &
5174 &    -7.2$\pm$0.7 &  185 &174 &13 &  5512 &  2.86 &\,& 5474 &  2.81  & 0.25   &$\le$15 \\    %   MEM  84 1418 3015 0.970 & 0.982 &
5176 &    -3.8$\pm$2.2 &  129 &117 &20 &  5393 &  2.38 &\,& 5336 &  2.32  & 0.29   &$\le$15 \\    %   MEM  91 1232 3597 1.008 & 1.027 &
5190 &    -8.6$\pm$0.6 &   23 &$\le$6&5&  5174 &$\le$0.67&\,& -- &   --   &  --    &$\le$15 \\    %   MEM  92 1443 --   1.083 & -9.99 &
5209 &    -7.3$\pm$0.7 &  122 &109 & 8 &  5336 &  2.27 &\,& 5294 &  2.23  & 0.22   &$\le$15 \\    %   MEM  41 1401 3088 1.027 & 1.041 &
5233 &    -9.2$\pm$0.7 &  146 &132 & 7 &  5066 &  2.10 &\,& 5103 &  2.14  & 0.22   &$\le$15 \\    %   MEM  30 1164  387 1.121 & 1.108 &
5235 &    -9.1$\pm$0.7 &  158 &143 & 8 &  4985 &  2.06 &\,& 4918 &  1.98  & 0.23   &$\le$15 \\    %   MEM  62 1224   26 1.149 & 1.172 &
5241 &    -9.4$\pm$0.6 &  161 &147 &20 &  5120 &  2.25 &\,& 5066 &  2.18  & 0.29   &$\le$12 \\    %   MEM  18 1111   55 1.102 & 1.121 &
5257 &    -7.8$\pm$0.6 &  131 &117 & 7 &  5174 &  2.14 &\,& 4982 &  1.91  & 0.21   &$\le$15 \\    %   MEM  26 1194   11 1.083 & 1.150 &
5263 &    -8.3$\pm$1.5 &  161 &147 &10 &  5094 &  2.22 &\,& 5020 &  2.13  & 0.25   &    --  \\    %   MEM  99 1149 3843 1.111 & 1.137 &
5267 &    -8.9$\pm$0.6 &  170 &155 &15 &  5011 &  2.16 &\,& --   &   --   & 0.27   &$\le$15 \\    %   MEM  23 1231 --   1.140 & -9.99 & 
5274 &    -8.8$\pm$0.8 &  181 &166 & 8 &  5040 &  2.26 &\,& 4897 &  2.08  & 0.24   &$\le$15 \\    %   MEM  29 1052 4093 1.130 & 1.179 &
5288 &    -7.8$\pm$0.5 &  130 &115 &10 &  5011 &  1.93 &\,& 4815 &  1.68  & 0.25   &$\le$15 \\    %   MEM  74 1076 4040 1.140 & 1.207 &
5289 &    -8.8$\pm$0.8 &  147 &133 &10 &  5066 &  2.10 &\,& 4772 &  1.73  & 0.24   &$\le$20 \\    %   MEM   9 1072 4050 1.121 & 1.222 &
5293 &    -8.1$\pm$0.8 &  151 &135 &13 &  4847 &  1.84 &\,& 4769 &  1.74  & 0.26   &$\le$15 \\    %   MEM  37 1193 3743 1.196 & 1.223 &
5297 &    -9.2$\pm$0.4 &  182 &167 &10 &  4956 &  2.16 &\,&  --  &   --   & 0.25   &$\le$15 \\    %   MEM  80 1341 --   1.159 & -9.99 &
5316 &    -8.8$\pm$0.7 &  150 &135 &14 &  4930 &  1.94 &\,& 4809 &  1.79  & 0.27   &$\le$15 \\    %   MEM  61 1078 4024 1.168 & 1.209 &
5317 &    -6.9$\pm$1.2 &   98 & 82 &20 &  4792 &  1.44 &\,& 4880 &  1.55  & 0.34   &$\le$15 \\    %   MEM  25 1388 3145 1.215 & 1.185 & 
5332 &    -7.8$\pm$0.2 &  165 &147 &15 &  4658 &  1.65 &\,& 4446 &  1.36  & 0.29   &$\le$15 \\    %   MEM  47 1237 3591 1.262 & 1.348 &
5339 &    -8.5$\pm$0.5 &   82 & 65 &10 &  4683 &  1.16 &\,& 4647 &  1.11  & 0.29   &     12 \\    %   MEM   7 1141 3880 1.253 & 1.266 &
5345 &    -8.5$\pm$0.4 &  230 &211 &15 &  4391 &  1.65 &\,& 4348 &  1.59  & 0.29   &$\le$15 \\    %   MEM  43 1229 3623 1.375 & 1.398 &
5357 &    -8.1$\pm$1.0 &  124 &107 &20 &  4683 &  1.46 &\,& 4573 &  1.30  & 0.32   &$\le$20 \\    %   MEM  22 1303 3384 1.253 & 1.294 & 
5376 &   -10.8$\pm$1.0 &  -99 &$\le$4&4&  4468 &$\le$-0.43&\,& --&   --   &  --    &$\le$15 \\    %   MEM  35 1047 --   1.338 & -9.99 &
5388 &    -7.6$\pm$1.6 &  157 &138 &20 &  4391 &  1.23 &\,&  --  &   --   & 0.31   &    --  \\    %   MEM  97 1267 --   1.375 & -9.99 &
5407 &    -9.3$\pm$1.0 &   68 & 49 &12 &  4448 &  0.69 &\,&  --  &   --   & 0.32   &     12 \\    %   MEM  87 1133 --   1.347 & -9.99 &
5408 &    -7.2$\pm$4.1 &  311 &291 &25 &  4292 &  2.05 &\,&  --  &   --   & 0.34   &     42 \\    %   MEM  67 1407 --   1.432 & -9.99 &
5410 &    -6.2$\pm$1.4 &   86 & 67 &20 &  4429 &  0.83 &\,&  --  &   --   & 0.37   &    --  \\    %   MEM  98 1385 --   1.356 & -9.99 &
5419 &    -5.7$\pm$0.6 &   18 &$\le$6&6&  4264 &$\le$-0.55&\,& --&   --   &  --    &     12 \\    %   MEM  65 1291 --   1.451 & -9.99 &
5437 &    -6.6$\pm$0.7 &   29 &$\le$4&4&  4372 &$\le$-0.58&\,& --&   --   &  --    &$\le$15 \\    %   MEM  32 1099 --   1.385 & -9.99 &
\hline
\end{tabular}
$\,$\\   %
SB1999: Values derived using (V-I)c colors from Song \& Bessell (1999)  .\\
ByN2000: We used values from Barrado y Navascu\'es et al. (2000b) otherwise.\\
\end{table}

\setcounter{table}{2}

\begin{table}
\caption[]{Iron equivalent widths for the brightest members of M35, in m\AA.}
\begin{tabular}{crrrrrrrrr}
\hline
Fe{\sc I} $\lambda$$\lambda$  & \multicolumn{9}{c}{W(FeI) of M35 candidate members, identification numbers} \\
\cline{2-10}
   (\AA)     & \#5098  &\#5103  &\#5125   & \#5089  & \#5107  &\#5135   & \#5096   & \#5116   & \#5174  \\  % 

%    (1)      &    (2)  &   (3)  &   (4)   &    (5)  &    (6)  &   (7)   &    (8)   &    (9)   &    (10) \\  % 
%M35, star    &   1082  &  1119  &  1144   &   1169  &   1225  &  1254   &   1393   &   1309   &   1418  \\  %    
%
\hline     	       	                  	    	                          	                	     	    	    		   
  6481.870   &   53    &   --   &   --    &   --    &   --    &   --    &   --     &   --     &   --    \\  %  --      --                       
  6569.215   &   56    &   96   &   64    &   66    &   66    &   58    &   45     &   64     &   87    \\  %  2.279   0.1330E-02              
  6593.871   &   78    &  105   &   93    &   --    &   --    &   --    &   --     &   72     &  120    \\  %  4.733   0.3589E+00              
  6627.545   &   21    &   --   &   --    &   --    &   --    &   --    &   25     &   --     &   45    \\  %  2.433   0.5585E-02              
  6677.987   &  117    &  186   &  148    &  123    &  132    &  126    &  112     &  153     &  169    \\  %  4.548   0.3162E-01              
  6703.567   &   21    &   20   &   35    &   19    &   20    &   38    &   28     &   17     &   51    \\  %  2.692   0.5164E-01              
  6705.103   &   35    &   30   &   38    &   30    &   51    &   40    &   40     &   --     &   74    \\  %  2.758   0.9247E-03              
  6710.319   &    -    &   11   &   --    &   --    &   --    &   --    &   --     &   --     &   --    \\  %  4.607   0.9099E-01              
  6726.667   &    -    &   --   &   32    &   --    &   --    &   --    &   38     &   --     &   45    \\  %  --      --                      
  6732.065   &    -    &   --   &   --    &   --    &   --    &    4    &   --     &   --     &   --    \\  %  --      --                      
  6733.151   &   14    &   --   &   --    &   --    &   26    &   21    &   27     &    9     &   45    \\  %  --      --                      
  6750.152   &   74    &   69   &   58    &   72    &   75    &   71    &   51     &   67     &   92    \\  %  4.638   0.3483E-01              
  6752.707   &   30    &   40   &   --    &   40    &   35    &   39    &   29     &   44     &   48    \\  %  2.424   0.2786E-02              
  6810.263   &   --    &   36   &   38    &   29    &   49    &   --    &   41     &   42     &   58    \\  %  4.638   0.6095E-01              
  6820.372   &   24    &   26   &   29    &   19    &   --    &   --    &   --     &   --     &   31    \\  %  --      --                      
  6828.591   &   --    &   --   &   --    &   --    &   --    &   --    &   28     &   --     &   --    \\  %  4.638   0.7328E-01              
  6837.006   &   --    &   --   &   --    &    6    &   --    &   --    &   --     &   --     &   --    \\  %  --      --                      
\hline     	       	                  	    	                          	                	     	    	    		   
\end{tabular}
\end{table}

\setcounter{table}{3}

\begin{table}
\caption[]{Metallicity for the brightest members of M35.}
\begin{tabular}{ccccccc}
\hline
     & \multicolumn{2}{c}{phot} &\,& \multicolumn{2}{c}{synthesis} \\
\cline{2-3}\cline{5-6}
 ID  & Teff  &[Fe/H]            &\,& Teff &      [Fe/H]    \\    %   A(Li)syn                     A(Li)cog 	
 \#  & (K)   &                  &\,& (K)  &                \\  
 (1) &    (2)&  (3)             &\,&  (4) &         (5)    \\
\hline	             			      
5089 &  5925 & -0.15            &\,&  5805 & -0.30$\pm$0.18 \\  % & 2.76$\pm$0.18   &     32 & --        &  %   MEM  44 1169 --   
5096 &  5849 & -0.12            &\,&  6000 & -0.16$\pm$0.18 \\  % & 2.76$\pm$0.21   &$\le$15 & 2.96 0.10 &  %   MEM  83 1393 1030 
5098 &  5705 & -0.32            &\,&  5800 & -0.26$\pm$0.12 \\  % & 2.56$\pm$0.19   &     13 & 2.88 0.10 &  %   MEM  55 1082  178 
5103 &  5814 & -0.09            &\,&  5705 & -0.16$\pm$0.19 \\  % & 2.79$\pm$0.21   &     18 & 2.86 0.10 &  %   MEM  34 1119 --   
5107 &  5673 & -0.17            &\,&  5810 & -0.11$\pm$0.14 \\  % & 2.53$\pm$0.22   &     25 & 3.01 0.10 &  %   MEM  50 1225   32 
5116 &  5673 & -0.31            &\,&  5715 & -0.31$\pm$0.21 \\  % & 2.64$\pm$0.21   &     24 & 2.96 0.10 &  %   MEM   4 1309  251  
5125 &  5705 & -0.26            &\,&  5710 & -0.26$\pm$0.12 \\  % & 2.79$\pm$0.23   &     20 & 3.01 0.01 &  %   MEM   6 1144 3856 
5135 &  5480 & -0.33            &\,&  5815 & -0.14$\pm$0.17 \\  % & 2.53$\pm$0.22   &$\le$15 & 3.31 0.10 &  %   MEM  64 1254  128 
5174 &  5512 & -0.04            &\,&  5700 & +0.10$\pm$0.19 \\  % & 2.86$\pm$0.25   &$\le$15 & 3.51 0.10 &  %   MEM  84 1418 3015 
\hline				  
\end{tabular}			  
$\,$ \\
We have  adopted Log N(Fe/H)$_\odot$=7.52 dex.
\end{table}